\documentclass[12pt, a4paper]{article}

\usepackage{amsmath,amssymb,graphicx}
\usepackage[english]{babel}
\usepackage{bm,url}
\usepackage[a4paper,text={17cm,25.5cm},centering]{geometry}
\usepackage[compact,small]{titlesec}
\usepackage{lmodern}
\usepackage{comment}
\usepackage{natbib}
\usepackage{nicefrac}
\usepackage{booktabs}
\usepackage{multirow, footnote}
\usepackage{color}
\usepackage{lscape}
\usepackage[hidelinks]{hyperref}
\usepackage{authblk}
\usepackage{scrextend}
\usepackage[table]{xcolor}
\usepackage{longtable, caption, array} 
\usepackage{adjustbox}
\usepackage{setspace}
\usepackage{subcaption}
\usepackage{framed}
\usepackage{dcolumn}
\definecolor{shadecolor}{RGB}{180,180,180}
\usepackage{float}
\usepackage{pdflscape}
\usepackage{siunitx}
\usepackage{listings}
\usepackage{xcolor}

\newtheorem{question}{Question}

\newcolumntype{L}[1]{>{\raggedright\let\newline\\ \arraybackslash\hspace{0pt}}m{#1}}
\newcolumntype{C}[1]{>{\centering\let\newline\\ \arraybackslash\hspace{0pt}}m{#1}}
\newcolumntype{R}[1]{>{\raggedleft\let\newline\\ \arraybackslash\hspace{0pt}}m{#1}}

\usepackage{scalerel,stackengine}
\stackMath
\newcommand\reallywidehat[1]{%
\savestack{\tmpbox}{\stretchto{%
  \scaleto{%
    \scalerel*[\widthof{\ensuremath{#1}}]{\kern-.6pt\bigwedge\kern-.6pt}%
    {\rule[-\textheight/2]{1ex}{\textheight}}%WIDTH-LIMITED BIG WEDGE
  }{\textheight}% 
}{0.5ex}}%
\stackon[1pt]{#1}{\tmpbox}%
}

\addto\extrasenglish{
  
}

\newcolumntype{T}{>{\footnotesize}c}
\begin{document}

\title{
The Connection Between Monetary Policy and Housing Prices: Public Perception and Expert Communication}

\date{ }
%\date{\today}

\author[a,c]{Philipp Poyntner}
\author[b,c]{Sofie R. Waltl\thanks{Corresponding Author}}
\affil[a]{University of Salzburg, Department of Economics, Rudolfskai 42, 5020 Salzburg, Austria; \url{philipp.poyntner@plus.ac.at}}
\affil[b]{University of Cambridge, Department of Land Economy, 17 Mill Lane, CB2 1RX Cambridge, UK; \url{sw2123@cam.ac.uk}}
\affil[c]{Vienna University of Economics and Business, Department of Economics\\ Welthandelsplatz 1, 1020 Vienna, Austria}

\clearpage\maketitle
\thispagestyle{empty}

\vspace{-1.5cm}
\begin{spacing}{0.9}

\begin{abstract}
We study how the general public perceives the link between monetary policy and housing markets. Using a large-scale, cross-country survey experiment in Austria, Germany, Italy, Sweden, and the United Kingdom, we examine households’ understanding of monetary policy, their beliefs about its impact on house prices, and how these beliefs respond to expert information. We find that while most respondents grasp the basic mechanisms of conventional monetary policy and recognize the connection between interest rates and house prices, literacy regarding unconventional monetary policy is very low. Beliefs about the monetary policy-housing nexus are malleable and respond to information, particularly when it is provided by academic economists rather than central bankers. Monetary policy literacy is strongly related to education, gender, age, and experience in housing and mortgage markets. Our results highlight the central role of housing in how households interpret monetary policy and point to the importance of credible and inclusive communication strategies for effective policy transmission.
\end{abstract}

\begin{footnotesize}
\textbf{Keywords:} Monetary Policy Communication, Experts, Credibility, Housing Markets, Survey-Experiment\\
\textbf{JEL codes:} E52, E58, R31, D83, D84, G53
\end{footnotesize}

\vfill
\begin{footnotesize}
\textbf{Notes and Acknowledgements:} We thank Michaele Diehl for excellent research assistance. Further, this work benefits from comments and suggestions by Kilian Rieder as well as participants in the \emph{1st Workshop on Residential Housing Markets: Perceptions and Measurement} in Luxembourg, the \emph{3rd Workshop on Residential Housing Markets: A Market in Distress and Potential Solutions} in Vienna, the \emph{38th IARIW General Conference} in London, the \emph{SCEUS Young Scholars Workshop} at the University of Salzburg, the \emph{5th International Workshop on Rent Control} in Tarragona, and the \emph{1st Housing Policy Symposium} at WU Vienna.

This research benefits from funding by the \emph{Jubil\"{a}umsstiftung der Wirtschaftsuniversit\"{a}t Wien}, Grant No. 27001764 (H-WELL), and internal funding provided by the research institute \emph{INEQ} at the Vienna University of Economics and Business. A pilot study benefitted from funding of the \emph{FNR Luxembourg National Research Fund}, CORE Grant No. 3886 (ASSESS). The experimental part was approved by the \emph{LISER Ethics Committee} on 31 March 2021 and has been pre-registered as \cite{ASSESS}.
\end{footnotesize}

\end{spacing}

\newpage

\section{Introduction}\label{sec:intro}

Ever since the global financial crisis (GFC) hit, the connection between monetary policy (MP, from now on) actions and (price) movements in housing markets has been under scrutiny by both academics \citep{taylor2007housing, jorda2015betting, hedlund2017monetary} and policy-makers \citep{bernanke2010monetary}. A common reasoning of this nexus can be summarised as follows: Changes in interest rates are expected to affect agents' decision-making and, thus, generate movements in real estate prices. Specifically, a reduction in interest rates decreases the cost of borrowing, alleviates credit constraints and, by that, increases the demand for housing. As the supply of housing is fixed in the short run, only prices can adjust and, thus, increased demand for housing triggers a rise in house prices \citep{adams2010macroeconomic}. Such effects have been shown to be particularly large in countries with strongly developed mortgage markets and the common use of external financing for housing purchases \citep{calza2013housing}. In this study, we focus on countries with well-developed mortgage markets that exhibit a broad range of housing market structures: those with large rental sectors (Austria and Germany), countries with intermediate homeownership rates (Sweden and the United Kingdom), and a country where homeownership is overwhelmingly predominant (Italy).

Despite the importance of this nexus \citep{jorda2016great},
relatively little is known about how this relationship between MP and housing is \emph{perceived} by the public. This is not only interesting \emph{per se}, but also because previous research has found that house price expectations play an important role in driving actual house prices \citep{glaeser2017extrapolative}. There are only few studies empirically linking MP directly to future house price expectations. One such exception is \cite{binder2025central} who have recently shown how house price expectations respond to communication about interest rate changes. Their survey-experiment  reveals that information about the mortgage channel of MP affects consumers' house price expectations.

The literature on how communication shapes expectations in the context of monetary policy has grown rapidly \citep[see][for a survey]{blinder2008central}. Yet, research that specifically examines the link between monetary policy and \emph{house price inflation expectations} remains scarce. Moreover, central bank communication has predominantly been studied from the perspective of financial experts and market participants, while the perceptions and understanding of the general public have received far less attention \citep{haldane2018central}.

Further, policymakers are increasingly interested in monitoring and influencing expectations that accompany the increase in research on the formation of these expectations. Generally, economic dynamics depend on the interaction between the actual conduct of monetary policy and agents' understanding of it \citep{eusepi2010central}. Specifically, households' expectations about future MP might affect not only their beliefs but also their behaviour. \citet{carvalho2014people} show that US households are, at least in part, aware of the basic principles underlying the Fed's MP principles -- the Taylor rule.

 \citet{coibion2022monetary} investigate monetary policy communication using randomized control trials. They find that providing simple information about inflation reduces households' forecasts by about one percentage point in a low-inflation setting. This effect is smaller if the respondents are given information about MP meetings via a news article, casting doubt on the diffusion of central bankers' messages via conventional media outlets. In terms of the targeted consumers, the study predominantly reports quite homogeneous responses. Related research analyses the effects of forward guidance \citep{d2022managing}, information provision about inflation targets \citep{binder2018household}, or past inflation rates \citep{cavallo2017inflation}.

The present study contributes to these branches of literature by investigating if and, if so how, agents assume, on average, monetary policy actions to influence specific asset prices, namely aggregate changes in house prices. To elicit the perceived nexus, we designed and conducted an interactive, web-based survey-experiment. We recruited about 3,800 participants from Austria, Germany, Italy, Sweden and the UK ensuring representation of the working-age population and, by that, external validity. We systematically ask questions about the assumed ex-ante nexus between MP actions and changes in house prices in the context of very heterogeneous housing markets. Further, we study the influence of information provision on the adaptation of beliefs. We study the heterogeneity in updated beliefs by the type of expert providing such information: either a central bank economist or an academic economist. To minimize suspicion among survey participants, we cite experts' publicly available statements and provide links to the original sources.
We thus can also provide some recommendations with regard to which experts appear to be trusted more.

Our study provides one of the first insights into the perceived link between monetary policy actions and housing prices by the general public. We identify a number of insights:
First, when asked directly about the functioning of MP, more than half of the respondents appear to understand the basic principles of conventional MP (higher prices will lead to an increase in policy rates). For unconventional MP (quantitative easing), only one tenth of respondents select the correct answer, revealing a substantial information and knowledge gap. 
A fine-grained assessment of the interplay between monetary policy literacy and socio-economic characteristics shows that knowledge of conventional monetary policy (CMP) is often acquired over the life course, particularly through high-stakes experiences such as real estate transactions. In contrast, literacy in unconventional monetary policy (UMP) appears to remain specialised expert knowledge that is seldom acquired through life experience. Further, we find systematically lower CMP literacy among German and Italian participants.
Second, 75\% of respondents think that there is a connection between interest rates and housing prices. Specifically, survey participants overwhelmingly link interest rate decreases to increasing house prices, in line with economic theory. 
Third, a non-negligible number of participants correct an initially wrong assessment of the nexus between MP and house prices when presented with indicative information by academic or central bank economists. This corroborates the recent findings of \citet{binder2025central}. 

Fourth, we contribute to the yet very small body of evidence on heterogenous impact of expert commentaries on the public opinion. \cite{hirschman2014economists} discuss the general impact of economists' institutional positions on authority and shaping policy, and find that expert economists appear to receive high professional authority. \cite{dommett2019we} also rejects the `post-truth' argument in an assessment of survey data from the UK and continental Europe as the vast majority appeared to see a value of experts' knowledge and policy advise. Yet, the main finding from this review is that there is currently too little understanding of the public perception of different types of experts.
Such heterogeneity across different experts' opinions are more rarely studied: One notable exception is the study by
\cite{sievertsen2025female} that detects a gender gap in credibility within the group of academic economists publicly providing expertise and advise.
In the present study, we aim to contribute to this strand of literature by studying heterogeneity of the credibility of experts' opinions across academic and policy economists. 
We find that participants are more reactive to information provided by academic economists compared to central bankers. We do not detect heterogeneities in this effect related to participants' stated trust in institutions and perceived corruption, democratic participation, education attainment, or other dimensions.
This general result therefore suggests that the public may be more reluctant to acknowledge expertise by central bankers as compared to academic economists, with important implications for central bank communication. 

Fifth, we contribute to the general literature on financial literacy and its implications for decision-making. General levels of financial literacy and education attainment have been shown to strongly correlate with the quality of financial decision-making including the likelihood of mortgage stress \citep{hu2024financial}, high-cost borrowing \citep{lusardi2015debt} or inadequate financial planning \citep{lusardi2014economic}.
In this spirit, we assess the association between specifically \emph{monetary policy literacy} and acquired expertise via past active involvement in housing markets and particularly transactions. Own involvement in real estate transactions including a mortgage are a clear predictor of CMP literacy yet not UMP. This notion is in-line with findings by \cite{bucks2008borrowers} documenting that mortgage-holders all appear to have a basic understanding of their mortgage terms yet fail in large proportions concerning more complex product features and appear often unprepared for unfavourable future interest rate scenarios.

Further, we tests whether MP literacy is associated with similar socio-economic characteristics as other forms of financial and economic literacy.
A meta-study regarding general financial literacy by \cite{lusardi2011financial} finds that on average women reach lower levels of financial literacy than men, the young and the old are less financially literate than the middle-aged, and financial literacy increases with formal education attained. The gender effect in financial literacy has been replicated ever since then in further studies expanding the scope of analysis \citep[e.g.,][]{cupak2018decomposing}.
Largely in-line with these previous findings, we document a clear association between literacy in conventional and unconventional monetary policy and both, education and gender -- with men outperforming women. Further, literacy in CMP first increases with age, peaks at the age of typical home purchase, and stabilises in older age. However, for UMP no such experience learning curve is detected.
When interacting age with gender, the gendered effect in CMP literacy remains significant indicating that women, on average, also gain less expertise over their lifetime.

We detect further nuanced heterogeneities proxying stability and engagement: participants stating more pronounced fears of losing their jobs are associated with significantly lower levels of literacy and likewise people not having made use of their voting rights in the most recent election are also associated with lower levels of literacy. The latter suggests that future research on various dimensions of financial and economic literacy may benefit from taking into account more nuanced measures associated with class.  
Concerning differences across country, we find the most solid understanding among Austrians.

The remainder of this article is structured as follows: Section \ref{sec:data} presents the study design and the data collected. Section \ref{sec:mp} documents findings on monetary policy literacy. The main part of our analysis of monetary policy and housing perceptions is documented in Section \ref{sec:beliefs} and Section \ref{sec:conclusions} concludes.

%%%%%%%%%%%%%%%%%%%%%%%%
\section{Data}\label{sec:data}

%%%%%%%%%%%%%%%%%%%%%
The study was fielded in May 2023, jointly with a cross-country data-collection endeavour aimed at measuring tax preferences in Austria, Germany, Italy, Sweden, and the UK.%
\footnote{A pilot study was run in Luxembourg between February and May 2022 together with participants' recruited for the study \cite{lepinteur2025equal}. Ethical approval has been obtained by the LISER Ethics Committee on 31 March 2021, and the study has been pre-registered
as \cite{ASSESS}.}
Participants were recruited in collaboration with the company CINT--GapFish, which maintains subject panels used for market and opinion research in various countries (see \url{https://gapfish.com/} for details).

In each country, we recruited approximately 800 participants using stratified random sampling based on socio-economic characteristics so as to represent the respective working-age population and ensure external validity of our results. The stratification dimensions were age, gender, highest level of education attained, and region of residence (urban vs.\ rural).

Upon log-in via the CINT-GapFish platform, participants were redirected to a web-based survey created using \emph{LIONESS Lab} \citep{giamattei2020lioness} and hosted on a server of the \emph{Vienna University of Economics and Business}. Respondents completed the survey independently; both access and completion times were recorded. To ensure clean data, participants could neither pause the survey and return later nor complete it multiple times from the same IP address.

Participants were paid for completing the study via the standard CINT-GapFish payment system and received remuneration only for a fully completed questionnaire.

To accommodate linguistic diversity across countries, multiple language versions of the survey were made available. In general, the survey was offered in English, German, Swedish, and Italian. The experiment, originally designed in English, was translated -- and, to ensure consistency across languages, back-translated as suggested by \cite{roth1991bargaining}. We carried out these translations using a combination of native speakers and AI-based tools. All participants could choose among the four language options.

While the substantive content of the questionnaire was identical across countries, country-specific elements—such as the list of regions of residence, the national currency, and the name of the respective central bank—were adapted to each national context.

Our final sample comprises 781 respondents from Austria, 738 from Germany, 818 from Italy, 836 from Sweden, and 688 from the UK. \autoref{tab:summary_all} in the Appendix reports the socio-economic and demographic characteristics of the participants, which we later use both to identify heterogeneities and as control variables.

Apart from these standard socioeconomic questions, we also use two subjective measures that have been shown to explain current and future well-being and behaviour:
subjectively perceived risk of \textit{job loss} (measured as the stated expectation of job loss during the next 6 months on a 10-steps scale) and the subjectively assessed degree of \textit{life satisfaction} (measured on a 10-steps scale). Both subjective measures have been shown to correlate with previous experience and future outcomes: subjective job security is correlated with previous experiences of job loss and predicts future unemployment \citep{campbell2007job, dickerson2012fears, 17Jobinsecurityandsubjectivewellbeing} and self-reported life satisfaction correlates with other measures of material well-being such as household wealth and income \citep{headey2004effects}. Further, these two measures have been shown to capture similar aspects of what could subsumed as ``a successful life'' \citep{layard2014predicts}.
Hence, we use these predictors as control variables but also to assess heterogeneity across respondents.

We also measure \emph{voter turnout} and political engagement by asking respondents whether they voted in the last election for which they were eligible. We draw from evidence from political sciences having agued in favour of the \emph{resource model} linking voter turnout with time, money, and skills proxied via having a job, a high income, and a high socio-economic status \citep{brady1995beyond}.

The respondents are asked six questions about monetary policy and the housing market, with two to four answer options each. These variables will be discussed in detail when the results are presented below.

%%%%%%%%%%%%%%%%%%%%%%%%%%%%%%%%%%%%%%%%

\section{Monetary Policy Literacy}\label{sec:mp}

We start with testing for participants' general MP literacy in the presence of inflation-targeting central banks. We assess aggregate results as well as heterogeneities across several dimensions.

%%%%%%%
\subsection{Aggregate Understanding of Conventional and Unconventional Monetary Policy}\label{sec:aggregate_MP}
Easy understanding by the public has been named as a key advantage of inflation targeting \citep{mishkin2000inflation}. To understand whether this prerequisite for more complex connections is given, we separately test for understanding of \emph{conventional} and \emph{unconventional} MP.

\begin{figure}[h]
    \begin{center}
    \caption{Answers to Question 1 (CMP) and Question 2 (UMP)}
    \begin{subfigure}{.45\textwidth}
  \centering
            \includegraphics[width=1\textwidth]{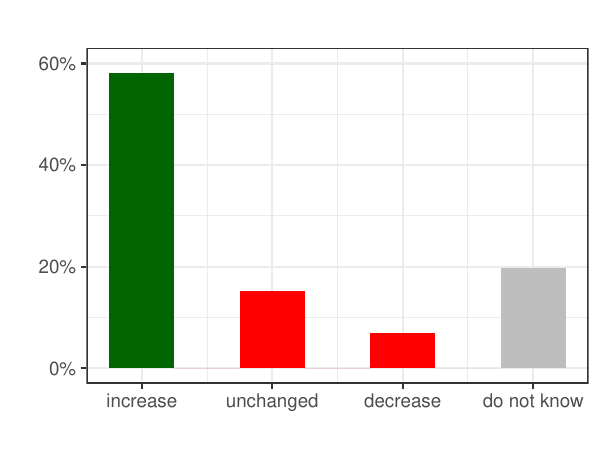}
            \caption{CMP}    
    \label{fig:fig_ECB_Q1_fr}
    \end{subfigure}\hfill
    \begin{subfigure}{.45\textwidth}
  \centering
    \includegraphics[width=1\textwidth]{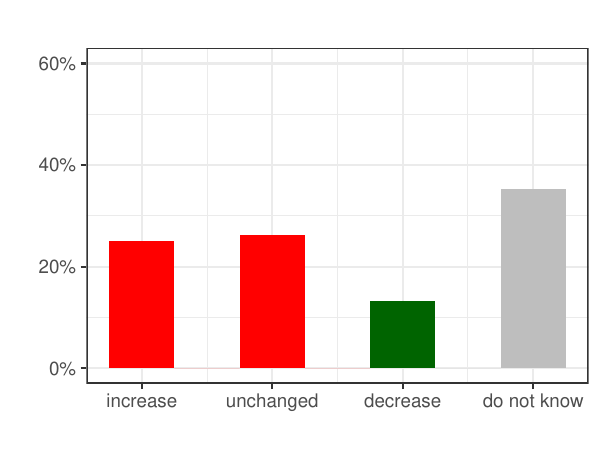}
        \caption{UMP}
    \label{fig:fig_ECB_Q2_fr}
    \end{subfigure}
    \begin{subfigure}{.45\textwidth}
    \centering
    \includegraphics[width=1\textwidth]{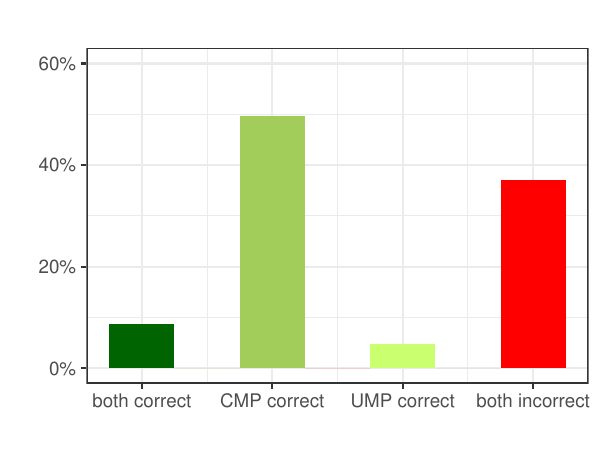}
        \caption{CMP $\cap$ UMP}
    \label{fig:fig_ECB_checker}
    \end{subfigure}

    \end{center}
    \begin{footnotesize}
        \emph{Notes:} The chart shows the distribution of answers to questions eliciting the connection between inflation and CMP (\autoref{fig:fig_ECB_Q1_fr}) and UMP (\autoref{fig:fig_ECB_Q2_fr}), respectively. The questions ask how interest rates (\autoref{fig:fig_ECB_Q1_fr}) or asset purchases (\autoref{fig:fig_ECB_Q2_fr}) are expected to change after an increase in inflation. See text for exact questions. \autoref{fig:fig_ECB_checker} shows the intersection of both questions. Correct answers are marked in green. Absolute counts are reported in \autoref{fig:replicate_Q1_Q2_counts} in the Appendix.
    \end{footnotesize}
\end{figure}

Question \ref{CMP.1}
asks about the general mechanism of conventional monetary policy (CMP) to understand whether respondents have a general knowledge about how interest rates emerge.%
\footnote{To avoid lengthy text in the questions, we exemplarily print ``ECB'' here whenever referring to the MP-conducting central bank. In the original questionnaires all text mentioning a central bank, the name of the institution is country-specific. Thus, for Austria, Germany and Italy the text refers to the ECB, while that for Sweden and the UK to the (Sveriges) Riksbank and the Bank of England, respectively.}

\begin{snugshade}
\begin{question}
(CMP)\label{CMP.1}\\The European Central Bank (ECB) together with national central banks decides upon the key policy interest rate, which affects the interest rates people have to pay when taking out mortgages.
  Suppose the prices in the Eurozone in general will go up in the next 12 months.
  How do you expect the ECB to react?
  \begin{enumerate}
    \item \textbf{The ECB will increase their interest rate.}
    \item The ECB will leave their interest rate unchanged.
    \item The ECB will decrease their interest rate.
    \item I do not know.
  \end{enumerate}
\end{question}
\end{snugshade}

As shown in \autoref{fig:fig_ECB_Q1_fr}, more than half of all respondents state that rising prices will trigger an increase in interest rates by the ECB.% 
\footnote{Throughout the paper, we indicate the answers that economists would most likely classify as correct by using boldface in the text and green bars in the figures.} %
About 22\% expect the ECB to act differently. \autoref{fig:fig_ECB_Q1_counts} in the Appendix shows the corresponding results in absolute counts.

\begin{figure}[h]
  \centering
\caption{Answers to Question 1 (CMP), by countries}\label{fig:Q1.countries}
  \begin{subfigure}[t]{0.32\textwidth}
    \centering
    \includegraphics[width=\linewidth]{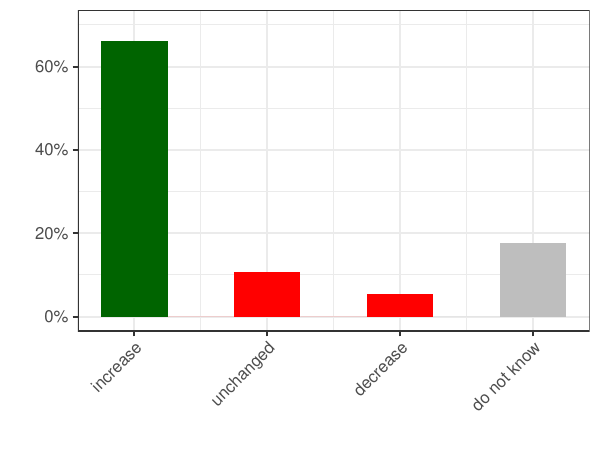}
    \caption{Austria}
    \label{fig:sub1}
  \end{subfigure}\hfill
  \begin{subfigure}[t]{0.32\textwidth}
    \centering
    \includegraphics[width=\linewidth]{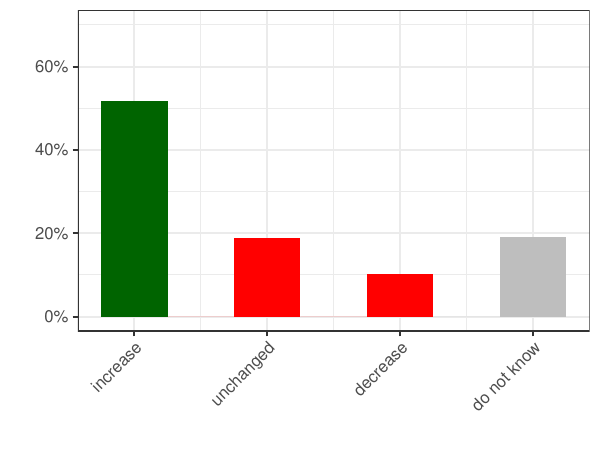}
    \caption{Germany}
    \label{fig:sub2}
  \end{subfigure}\hfill
  \begin{subfigure}[t]{0.32\textwidth}
    \centering
    \includegraphics[width=\linewidth]{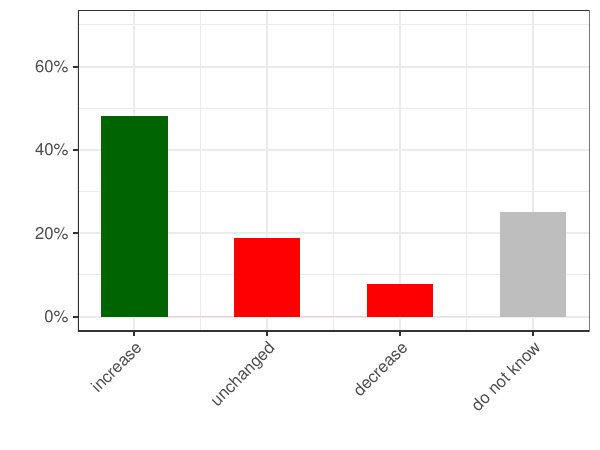}
    \caption{Italy}
    \label{fig:sub3}
  \end{subfigure}

  \vspace{0.8em}

  \begin{subfigure}[t]{0.32\textwidth}
    \centering
    \includegraphics[width=\linewidth]{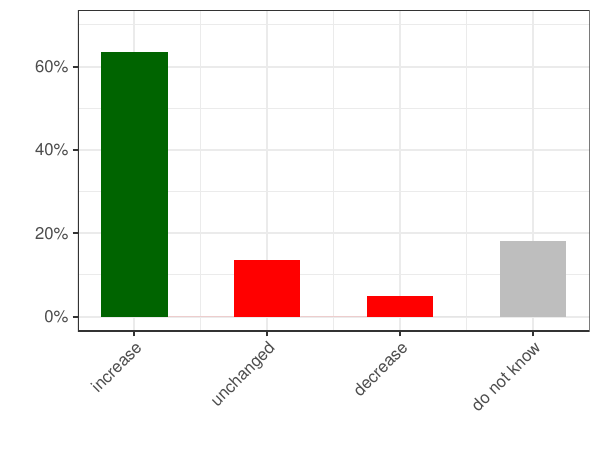}
    \caption{Sweden}
    \label{fig:sub4}
  \end{subfigure}\hfill
  \begin{subfigure}[t]{0.32\textwidth}
    \centering
    \includegraphics[width=\linewidth]{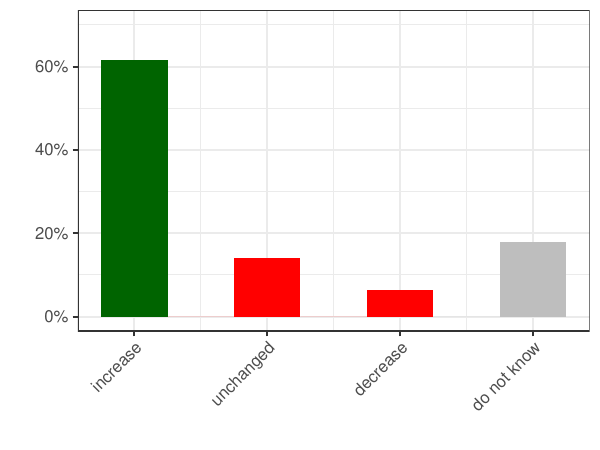}
    \caption{United Kingdom}
    \label{fig:sub5}
  \end{subfigure}  \hfill
\begin{subfigure}[t]{0.32\textwidth}
  \end{subfigure}
  \makebox[0.32\textwidth][c]{}
\begin{minipage}{0.95\textwidth}
\footnotesize{\emph{Notes:}
 The Figure shows country-specific distribution of understanding basic conventional monetary policy principles per country. Correct answers are highlighted in green. 
}
\end{minipage}
  \label{fig:fig_ECB_Q1_countries_fr}
\end{figure}

\clearpage

\begin{figure}[H]
    \centering
    \caption{Inflation, policy rates and house prices}
    \label{fig:inflationpolicy}
    
    \begin{subfigure}{\textwidth}
        \centering
        \caption{Inflation}
        \includegraphics[width=0.9\linewidth,height=0.25\textheight,keepaspectratio]{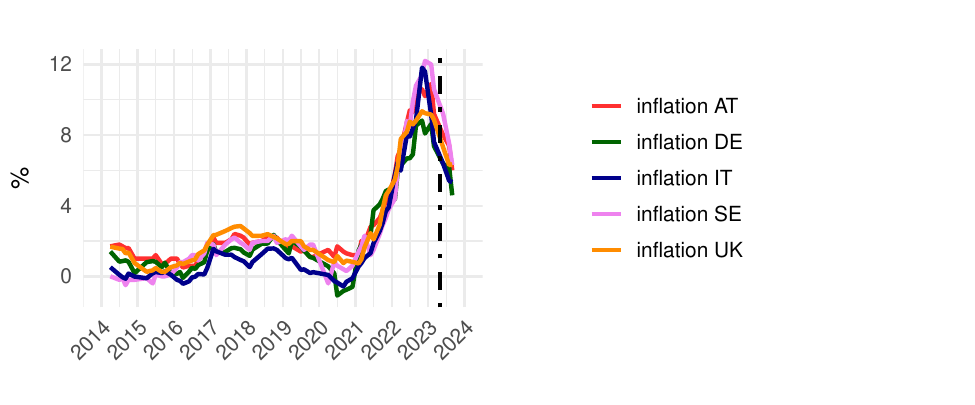}
        \label{fig:inflation_rates}
    \end{subfigure}
    
    \vspace{0.5em}
    \begin{subfigure}{\textwidth}
        \centering
        \caption{Policy and mortgage rates}
        \includegraphics[width=0.9\linewidth,height=0.25\textheight,keepaspectratio]{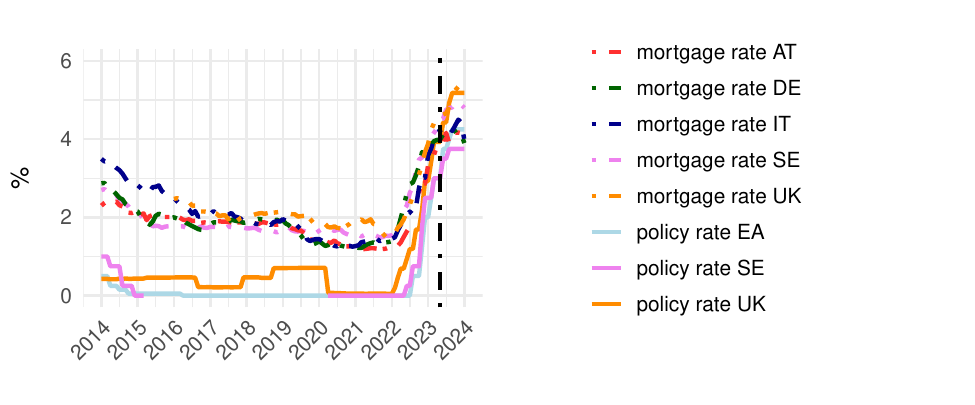}
        \label{fig:policyrates}
    \end{subfigure}
    
    \vspace{0.5em}
    \begin{subfigure}{\textwidth}
        \centering
        \caption{House prices (RPPIs)}
        \includegraphics[width=0.9\linewidth,height=0.25\textheight,keepaspectratio]{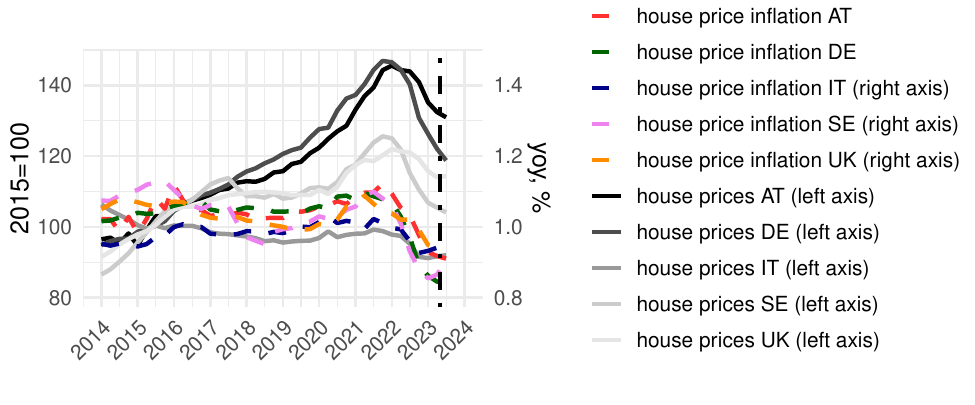}
        \label{fig:RPPIs}
    \end{subfigure}
    
    \caption*{\footnotesize \emph{Notes:} RPPI = ``Residential Property Price Indices''.\\
    \emph{Sources:} FRED, OECD, Statistics Sweden, OeNB, Bank of England, Banca d'Italia, Bundesbank.}
\end{figure}

\autoref{fig:Q1.countries} and
\autoref{tab:Q1_Q2} reveal some variation across countries: Respondents from Austria are most likely to expect an interest rate hike following a period of rising prices (`inflation'), while respondents from Germany and Italy do so less often.
Interestingly, all three countries have the same central bank (the ECB) setting a harmonised interest rate, and also very similar country-specific inflationary dynamics at the time of the survey as shown in \autoref{fig:inflationpolicy}. Such severe lack of basic MP literacy among German adults is inline with results from a survey-experiment also conducted during a high-inflation period \citep{drager2025inflation}.

We conduct a formal contingency $\chi^2$-test to test for country differences in survey responses. The test yields $\chi^2=93.658$ ($p$-value $<0.01$), which leads us to reject the null hypothesis of the answer distributions being the same across all countries. 
This indicates that countries with higher proportions of incorrect answers would benefit from improved education and clearer information on basic mechanisms.

Most striking about these knowledge gaps with regard to conventional MP is the timing: the survey was conducted in May 2023, and thus amidst a period of high and rapidly increasing inflation and, accordingly, substantial MP actions \citep[see again][]{drager2025inflation}. This was the case in all countries studied here as shown in \autoref{fig:inflationpolicy}. This finding is alarming, and raising public awareness of the measures taken to cushion inflationary pressures -- as well as the mechanisms behind them -- will be essential for maintaining public support.

\begin{table} 
\begin{center}    
  \caption{Answers to Question 1 (CMP) and 2 (UMP), by countries} 
  \label{tab:Q1_Q2} 
\begin{tabular}{@{\extracolsep{5pt}} l cccc c cccc} 
\toprule
\toprule
  & Question 1 (CMP) & n & Share && Question 2 (UMP) & n & Share\\ 
\cmidrule{2-4} \cmidrule{6-8} 
Austria & \textbf{increase} & $515$ & $0.659$ && purchase more    & $215$ & $0.275$ \\ 
        &  unchanged & $84$ & $0.108$         && no change        & $200$ & $0.256$ \\ 
        & decrease & $42$ & $0.054$           && \textbf{purchase less} & $113$ & $0.145$\\ 
        & do not know & $137$ & $0.175$       && do not know      & $252$ & $0.323$ \\ 
        & NA & $3$ & $0.004$                  && NA               & $1$ & $0.001$ 
\\ 
\cmidrule{2-4} \cmidrule{6-8} 

Germany & \textbf{increase} & $370$ & $0.501$ && purchase more    & $178$ & $0.241$ \\
        &  unchanged & $134$ & $0.182$        && no change        & $200$ & $0.271$ \\ 
        & decrease & $73$ & $0.099$           && \textbf{purchase less}  & $91$ & $0.123$ \\ 
        & do not know & $137$ & $0.186$       && do not know      &  $243$ & $0.329$ \\ 
        & NA & $24$ & $0.033$                  && NA              & $26$ & $0.035$
\\ 
\cmidrule{2-4} \cmidrule{6-8} 
Italy & \textbf{increase} & $387$ & $0.473$ && purchase more    & $208$ & $0.254$ \\ 
      &  unchanged & $152$ & $0.186$ && no change        & $223$ & $0.273$ \\ 
      & decrease & $62$ & $0.076$  && \textbf{purchase less}  & $129$ & $0.158$ \\ 
      & do not know & $202$ & $0.247$  && do not know      & $249$ & $0.304$ \\ 
      & NA & $15$ & $0.018$ && NA              & $9$ & $0.011$ \\ 
\cmidrule{2-4} \cmidrule{6-8}  
Sweden & \textbf{increase} & $524$ & $0.627$ && purchase more     & $168$ & $0.201$ \\ 
       &  unchanged & $113$ & $0.135$ && no change        & $183$ & $0.219$ \\ 
       & decrease & $40$ & $0.048$  && \textbf{purchase less} & $109$ & $0.130$ \\ 
       & do not know & $149$ & $0.178$  && do not know      &   $367$ & $0.439$ \\ 
       & NA & $10$ & $0.012$ && NA              & $9$ & $0.011$ \\ 
\cmidrule{2-4} \cmidrule{6-8} 
UK & \textbf{increase} & $389$ & $0.565$ && purchase more    & $171$ & $0.249$ \\ 
   &  unchanged & $89$ & $0.129$ && no change        & $183$ & $0.266$ \\ 
   & decrease & $41$ & $0.060$  && \textbf{purchase less} & $59$ & $0.086$ \\ 
   & do not know & $113$ & $0.164$  && do not know      & $218$ & $0.317$ \\ 
   & NA & $56$ & $0.081$ && NA              & $57$ & $0.083$ \\ 
%  & NA & $2$ & $1$ \\ 
\bottomrule
\bottomrule
\end{tabular} 
\end{center}
\begin{footnotesize}
    \emph{Notes:} The Table reports answers to Question 1 (Conventional Monetary Policy) and Question 2 (Unconventional Monetary Policy). Correct answers are emphasised. 
\end{footnotesize}
\end{table}

Next, question \ref{UMP} tests for the basic understanding of unconventional monetary policy (UMP).

\begin{snugshade*}
\begin{question}
(UMP)\label{UMP}\\
Besides fixing the interest rate, the ECB also bought a range of financial assets, including government bonds and corporate bonds. Such purchases influence broader financial conditions and, possibly, inflation. Assume, again, that prices in the Eurozone in general increase over the next 12 months. How do you think the ECB will react?
 \begin{enumerate}
    \item The ECB will purchase more assets.
    \item The ECB will not change its asset purchase programme.
    \item \textbf{The ECB will purchase less assets.}
    \item I do not know.
\end{enumerate}
\end{question}
\end{snugshade*}

\autoref{fig:fig_ECB_Q2_fr} reveals that UMP is much less understood than CMP. 
More than a third of respondents explicitly state not to know how the ECB would react, meaning a significant share of participants is aware of their knowledge gaps. About half of the respondents incorrectly think that the ECB would increase asset purchases or leave the amount unchanged. Only 13\% -- thus a minority -- correctly expect a decrease in asset purchases. The remaining 87\% either state that they do not know the correct answer or tick a wrong answer.

\begin{figure}[htbp]
  \centering
\caption{Answers to Question 2 (UMP), by countries}
  \begin{subfigure}[t]{0.32\textwidth}
    \centering
    \includegraphics[width=\linewidth]{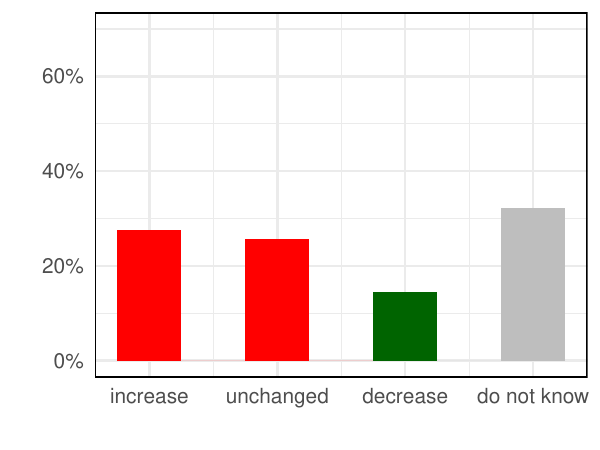}
    \caption{Austria}
    \label{fig:sub1}
  \end{subfigure}\hfill
  \begin{subfigure}[t]{0.32\textwidth}
    \centering
    \includegraphics[width=\linewidth]{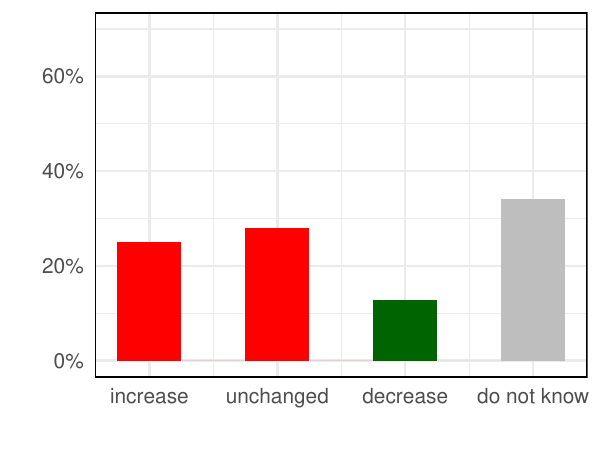}
    \caption{Germany}
    \label{fig:sub2}
  \end{subfigure}\hfill
  \begin{subfigure}[t]{0.32\textwidth}
    \centering
    \includegraphics[width=\linewidth]{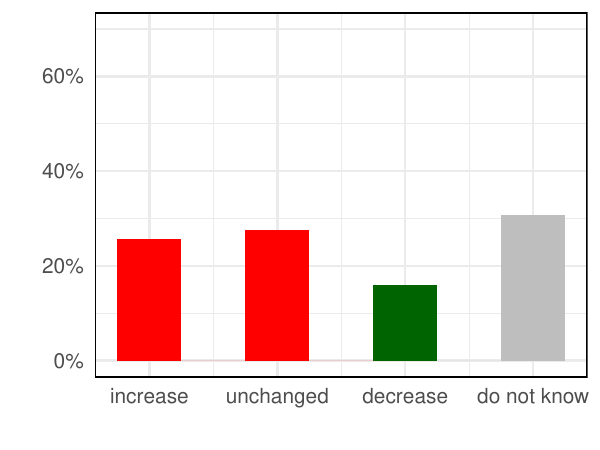}
    \caption{Italy}
    \label{fig:sub3}
  \end{subfigure}

  \vspace{0.8em}

  \begin{subfigure}[t]{0.32\textwidth}
    \centering
    \includegraphics[width=\linewidth]{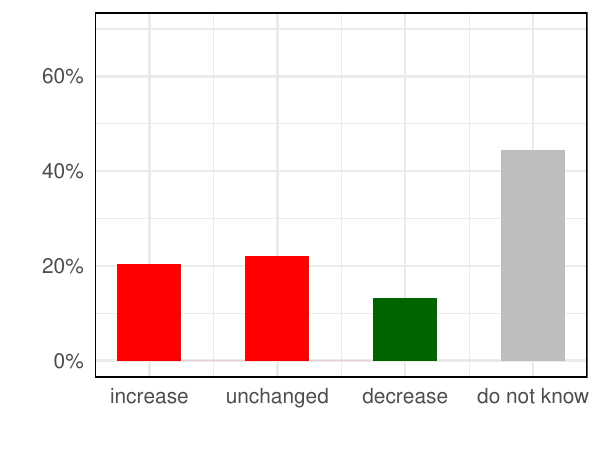}
    \caption{Sweden}
    \label{fig:sub4}
  \end{subfigure}\hfill
  \begin{subfigure}[t]{0.32\textwidth}
    \centering
    \includegraphics[width=\linewidth]{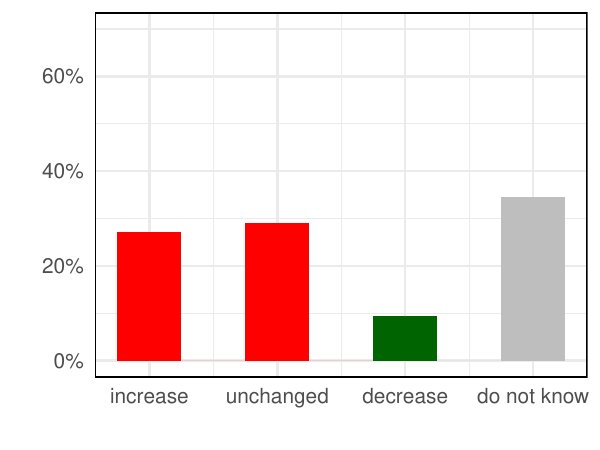}
    \caption{United Kingdom}
    \label{fig:sub5}
  \end{subfigure}  \hfill
\begin{subfigure}[t]{0.32\textwidth}
  \end{subfigure}
  \makebox[0.32\textwidth][c]{}
\begin{minipage}{0.95\textwidth}
\footnotesize{\emph{Notes:}
 The Figure shows country-specific distribution of understanding basic unconventional monetary policy principles per country. Correct answers are highlighted in green. 
}
\end{minipage}
  \label{fig:fig_ECB_Q2_countries_fr}
\end{figure}

\autoref{fig:fig_ECB_Q2_countries_fr} decomposes aggregate results into country-specific ones. The response ``I do not know'' is the most common answer selected in \emph{every single country}. Adding up stated and revealed knowledge gaps shows that the vast majority of participants (roughly 87\%) does not understand UMP. 
We conduct again a formal contingency $\chi^2$-test to test for country-differences, which yields a $\chi^2$ of 58.206 with a $p$-value$<0.01$, thus rejecting the null of answer distributions being the same for all countries. 
This hints toward a lack of effective MP communication and education.

We conclude that both revealed and stated knowledge of UMP is quite limited and this is true across all five countries assessed. 

But why is UMP so much less understood than CMP? The poor understanding of UMP could simply stem from the fact that this form of monetary policy constitutes a relatively recent policy tool: formal inflation targeting (CMP) was pioneered by New Zealand in the early 1990s \citep[see][]{haldane1995targeting} and has been adopted ever since then in all advanced economies globally. In contrast, asset purchase programs and other forms of UMP have only seen widespread application in the aftermath of the GFC in 2007/2008%
\footnote{With the notable exception of Japan, where quantitative easing has been undertaken since 2001.} \citep{bernanke1997inflation, werner1995create}. Whatever the reason, monetary policy actions demand public support and understanding. Thus, more advanced understanding would be needed and further justifies such educational initiatives run by central banks.%
\footnote{See \url{https://www.ecb.europa.eu/ecb-and-you/financial_literacy_europe/html/index.en.html} for an overview of public education initiatives in the Euro Area.}

Finally, we jointly assess responses to the CMP and UMP questions.
Overall, we find that only about 9\% have answered correctly to both of them and can thus be labelled ``MP experts''  (\autoref{fig:fig_ECB_checker}). While there are no large differences across countries, according to our study, the shares of ``MP experts' is highest in Austria (10\%) and with 6\% lowest in the UK.
Looking to the other extreme, overall 15\% give wrong answers on both questions (highest: Italy and Germany 18\%, lowest: Austria 11\%). 

A large share of participants answer with ``I do not know'' on both questions indicating also awareness of knowledge gaps: 15\% overall, whereas the share is largest in Italy (20\%) and lowest in Austria (13\%). 
However, a pairwise Pearson's $\chi^2$ test rejects the null hypothesis of independence between the two outcomes at the $p<0.01$-level ($\chi^2 = 9.5504$, $p\text{-value} = 0.001999$), yet the association is rather weak, as indicated by Cramér's $V = 0.051$%$%
\footnote{When performing these tests at the country level, we find positive associations for Germany and Italy, but none for the other countries; see \autoref{tab:chi_cramer_country}.}
suggesting that knowledge gaps in both concepts often, though not consistently, occur together.

%%%%%%%%%%%%%%%%
\subsection{Heterogeneity in Monetary Policy Literacy}\label{sec:hetero_MP}
We explore the association between MP literacy and socioeconomic characteristics to detect which part of the population would be in need of additional information. 
To that end, we estimate two logit regression models -- one for each type of MP -- with an indicator as outcome variable that takes the value of one for correct answers and zero otherwise. 
\autoref{tab:hetero_CMP} and \autoref{tab:hetero_UMP} report results.
We can also use this setup to detect associations between MP literacy and several relevant other dimensions.
First, model (1) adds respondents' socio-economic characteristics to the model.%
\footnote{The model reported here includes only variables that have coefficients significantly different from zero. \autoref{tab:hetero_MP_all} in the Appendix reports a regression model that includes other variables as well.} 
As a plausibility check, we compare the estimated intercepts in these two models to the univariate assessment in \autoref{sec:mp}: the intercept associated with the UMP model is substantially smaller than the one for CMP reiterating that, on average, people are more familiar with CMP than with UMP. 

Further, we assess whether MP literacy is associated with similar socio-economic characteristics as other types of financial and economic literacy.
In-line with the bulk of previous findings surveyed by \cite{lusardi2011financial}, we find that on average men have a better understanding of both conventional and unconventional MP. Even when interacting the gender dummy with education attainment in \autoref{tab:hetero_MP_all}, the positive effect for men remains significant and similar in size, indicating that gender does not measure hidden discrepancies in education attainment between men and women, but truly a difference in MP literacy. 

With regard to age, we would expect a diminishing marginal effect, which we would interpret as a learning process reaching its life-long maximum when people reach the typical age of first home purchase, and a stabilization (or even decline) of MP literacy thereafter.
While we indeed detect such a profile in correct responses to the CMP question, we find no age effect at all for the UMP question (see \autoref{tab:age} for a test of the functional form of the age effect). This again indicates that literacy in unconventional monetary policy is not general knowledge acquired over the life course, but instead remains a form of specialized expertise.

Non-surprisingly, variation in education levels%
\footnote{We distinguish between below and above tertiary education.} attained also significantly explain variation in correct answers: having attained tertiary education is associated with a roughly 30\% higher likelihood to tick the right answer for both CMP and UMP questions.

Examining additional characteristics, we find that individuals who are more likely to expect job loss within the next six months exhibit a poorer understanding of conventional monetary policy, consistent with the idea that greater job insecurity proxies for lower levels of education and economic status (see \autoref{sec:data} for references).

Thus, we can conclude that prime socio-economic variables explaining heterogeneity in general financial literacy are also present in MP literacy. This is particularly the case for CMP.

We further report some differences across countries: In comparison to Austrian respondents (the baseline in our regression), all others are less likely to expect interest rate hikes following price increases. Yet, in terms of magnitude there are differences across the remaining countries: people in Germany and Italy are much less likely to expect the correct CMP reaction. Regarding asset purchase programmes, UK respondents are least likely to give the correct answer.
Lastly, respondents that reported to not have voted in the last election are less likely to associate inflation with interest hikes. 
The following variables have no explanatory power for giving correct answers: marital status, having children, employment status, earnings, rural/urban, class (self-evaluated), and life satisfaction.

Overall, we detect knowledge gaps with regards to CMP among less educated participants, those in a perceived less secure labour market situation, and those who did not participate in democratic processes by voting (though they were eligible). Regarding UMP, we only identify a significant difference by gender.

Second, in model (2), we include variables that capture expectations and home-ownership. 
We elicit two types of expectations and test ex post whether they materialized. If participants stated expectations are close to realized developments, we consider this as a hint towards sound financial and economic literacy.
The first type of expectations refers to property price changes in \% over the upcoming 12 months. Participants are asked to allocate their expectations to one of five bins: $<-$5\%, $-$5\% to $-$2\%, $-$2\% to 2\%, 2\% to 5\%, over 5\%. In hindsight, we can turn to real data to see whether their predictions materialised  (``future house prices correct''=1), or not (``future house prices correct''=0). This forecasting ability is not correlated with correct monetary policy assessments of both types.
The second type of expectations alludes to real income growth. We find that people who expect their own income to increase less than general prices (a loss in real terms, which was indeed the case over the study period)%
\footnote{``Over the next 10 years, do you expect your (household’s) total income to go up more than prices, less than prices, or about the same as prices?''} %
are on average more likely to correctly answer the CMP question. There is no connection with UMP literacy detected.

Assessing CMP literacy by ownership status, we find that owner-occupiers with mortgages exhibit the highest likelihood to respond correctly in comparison to current renters and owners without mortgages.
These differences are only statistically significant for the CMP question yet not the UMP question.
That might indicate that people with an active mortgages may monitor factors that influence their mortgage payments, such as interests set by central banks (CMP) than others, and by that have gained a better understanding of these relationships.
There is no such effect for UMP, suggesting that asset purchase programs are on average not perceived to be directly associated with mortgages.

Third, models (3) to (5) include variables that reflect respondents' past experiences and future plans in the domain of housing markets.  
In model (3), we restrict the sample to owner-occupiers and differentiate them by mode of home acquisition. Those who purchased their homes (baseline), i.e., went through an ordinary transaction process themselves, are associated with the highest likelihood of correctly answering the CMP question. Those who have not been involved in a purchase as they have inherited their property or have received it as a gift, fare worse. 
Again, such connections are not detected for UMP literacy and, thus, together with the results from model (3) this indicates that purchasing a house, i.e., being involved in a major financial transaction, is not associated with a better understanding of UMP -- even when this purchase involves a mortgage. This finding is alarming given the high stakes involved.

Similarly, models (4) and (5) restrict the sample to owner-occupiers (4) or renters (5), respectively. Both are asked whether they want to change their tenure status in the future. Renters that plan to become owners in the future are associated with a slightly higher likelihood to correctly answer the CMP question yet the difference is not statistically significant from zero. In contrast, the restricted model for current owners detects a significant difference in knowledge between those aiming to stay owners in the future versus those that can imagine to rent in the future. The likelihood of a correct answers is as expected higher for prospective all-time-owners. Again, there are no such effects for UMP.

Overall, we can characterise people that do not well understand CMP as young, female, low-educated and less engaged in political participation. Further, they are on average more likely to fear job loss and most likely reside -- among the countries part of our study -- in Germany or Italy and least likely in Austria.

The profile of respondents selecting incorrect answers to the UMP question is similar: the least informed individuals are, on average, more likely to be female and have lower levels of education, and they are most likely to reside in the UK. By contrast, age, perceived job loss risk, and political participation do not appear to play a significant role. \autoref{tab:hetero_MP_all} in the Appendix reports the regression results when all covariates are included simultaneously; conclusions are robust.

As our country-samples are not exactly of the same size due to differential non-response behaviour, we re-estimate these models introducing a weighting scheme that enforces equal weight per country. Results are almost unchanged as reported in \autoref{tab:hetero_CMPw} and \autoref{tab:hetero_UMPw} in the Appendix supporting the robustness of our findings.

\begin{table}[!htbp] 
  \caption{Determinants of CMP Literacy} 
  \label{tab:hetero_CMP} 
  \begin{center}
  \resizebox{0.9\textwidth}{!}{    
  
\begin{tabular}{@{\extracolsep{5pt}}lccccc} 
\toprule \toprule
 & \multicolumn{5}{c}{\textit{Dependent variable: CMP correct}} \\ 
\cmidrule{2-6} 
\\ & \multicolumn{1}{c}{SE} &\multicolumn{1}{c}{Expec} & \multicolumn{3}{c}{Homeownership
}   \\ 

\\ & (1) & (2) & (3) & (4) & (5) \\ 
\midrule
Gender (Ref.: female)\\
 $\quad$ male & 0.246$^{***}$ &  &  &  &   \\ 
  & (0.069) &  &  &  &   \\ 
  Education attainment (Ref.: low)\\
   $\quad$ high education & 0.280$^{***}$ &  &  &  &    \\ 
  & (0.076) &  &  &  &  \\ 
  Age & 0.014$^{***}$ &  &  &  &  \\ 
  & (0.003) &  &  &  &   \\ 
  Likelihood of job loss within 6 months (10 steps Likert scale) & $-$0.073$^{***}$ &  &  &  &   \\ 
  & (0.013) &  &  &  &  \\ 
  Country Sample (Ref.: Austria)\\
    $\quad$ Germany & $-$0.697$^{***}$ &  &  &  &  \\ 
  & (0.110) &  &  &  &  \\ 
  $\quad$ Italy & $-$0.671$^{***}$ &  &  &  &  \\ 
  & (0.107) &  &  &  &  \\ 
 $\quad$ Sweden & $-$0.239$^{**}$ &  &  &  & \\ 
  & (0.108) &  &  &  &  \\ 
 $\quad$ United Kingdom & $-$0.305$^{***}$ &  &  &  &  \\ 
  & (0.116) &  &  &  &  \\ 
  Did not vote  (Ref. did vote)& $-$0.414$^{***}$ &  &  &  &  \\ 
  & (0.109) &  &  &  &   \\ 
  Future house prices correctly predicted (Ref. incorrect) &  & 0.010 &  &  & \\ 
  &  & (0.073) &  &  &   \\ 
  Expects that own income will over the next 10years...\\
  $\quad$ increase less than prices  (Ref. more or equal)&  & 0.349$^{***}$ &  &  &   \\ 
  &  & (0.090) &  &  &   \\ 
  $\quad$ increase the same as prices  &  & 0.009 &  &  &   \\ 
  &  & (0.103) &  &  &   \\ 
  Tenure status (Ref. renter)\\
   $\quad$ own house w/o mortgage  &  & 0.113 &  &  &  \\ 
  &  & (0.076) &  &  &   \\ 
   $\quad$ own house with mortgage &  & 0.475$^{***}$ &  &  &  \\ 
  &  & (0.090) &  &  &   \\
  Acquisition mode (Ref.: purchased)\\ 
   $\quad$ house gifted  &  &  & $-$0.241 &  &  \\ 
  &  &  & (0.174) &  &  \\ 
   $\quad$ house inherited  &  &  & $-$0.318$^{**}$ &  &   \\ 
  &  &  & (0.140) &  & \\ 
   $\quad$ other  &  &  & 0.116 &  &  \\ 
  &  &  & (0.218) &  &  \\ 
  Owns other real estate = yes &  &  & $-$0.065 &  &  \\ 
  &  &  & (0.104) &  &  \\ 
  Was involved in housing transactions in the past = No  &  &  & $-$0.422$^{***}$ &  &   \\ 
  &  &  & (0.091) &  &   \\ 
  Future plans (Ref.: renter wanting to become an owner)\\
  Is renter, does not want to become home owner  &  &  &  & $-$0.057 &   \\ 
  &  &  &  & (0.109) &    \\ 
    Future plans (Ref.: owner wanting to become a renter)\\
  is home owner, does not want to become home renter &  &  &  &  & 0.228$^{**}$   \\ 
  &  &  &  &  & (0.114)   \\ 
  Constant & 0.563$^{***}$ & $-$0.021 & 0.699$^{***}$ & 0.218$^{**}$ & 0.233$^{**}$  \\ 
  & (0.201) & (0.090) & (0.069) & (0.085) & (0.105)  \\ 
 \midrule
Observations & 3,729 & 3,702 & 2,252 & 1,433 & 2,310  \\ 
Log Likelihood & $-$2,436.608 & $-$2,487.670 & $-$1,491.602 & $-$987.144 & $-$1,548.543  \\ 
AIC & 4,893.216 & 4,987.341 & 2,995.205 & 1,978.289 & 3,101.086 \\ 
\bottomrule
\bottomrule
\end{tabular}} 
\end{center}
\begin{footnotesize}
    \emph{Notes:} The Table reports estimation results for probit models regressing the probability of a correct answer to Question 1 (CMP) on sets of control variables: respondents' socio-economic characteristics, general financial literacy, and previous housing market experience. $^{*}$p$<$0.1; $^{**}$p$<$0.05; $^{***}$p$<$0.01. 
\end{footnotesize}

\end{table} 

\begin{table}[!htbp]  
  \caption{Determinants of UMP Literacy} 
  \label{tab:hetero_UMP} 
  \begin{center}
  \resizebox{0.9\textwidth}{!}{    
\begin{tabular}{@{\extracolsep{5pt}}lccccc} 
\toprule
\toprule
 & \multicolumn{5}{c}{\textit{Dependent variable: UMP correct}} \\ 
\cmidrule{2-6} 
& \multicolumn{1}{c}{SE} & \multicolumn{1}{c}{Expec} & \multicolumn{3}{c}{Homewonership} \\ 
\\ & (1) & (2) & (3) & (4) & (5) \\ 
\midrule \\ 
Gender (Ref.: female)\\
 $\quad$ male & 0.423$^{***}$ &  &  &  &  \\ 
  & (0.099) &  &  &  &   \\ 
Education attainment (Ref.: low)\\
 $\quad$ high education & 0.328$^{***}$ &  &  &  &   \\ 
  & (0.106) &  &  &  &  \\ 
Age & $-$0.001 &  &  &  &   \\ 
  & (0.004) &  &  &  &  \\ 
Likelihood of job loss within 6 months (10steps Likert scale) & $-$0.002 &  &  &  &  \\ 
  & (0.018) &  &  &  &  \\ 
Country Sample (Ref.: Austria)\\
 $\quad$ Germany & $-$0.196 &  &  &  &   \\ 
  & (0.154) &  &  &  &  \\ 
 $\quad$ Italy & 0.130 &  &  &  &  \\ 
  & (0.144) &  &  &  &  \\ 
 $\quad$ Sweden & $-$0.169 &  &  &  &  \\ 
  & (0.148) &  &  &  &   \\ 
 $\quad$ United Kingdom & $-$0.589$^{***}$ &  &  &  &   \\ 
  & (0.175) &  &  &  &  \\ 
Did not vote  (Ref. did vote) & 0.064 &  &  &  &   \\ 
  & (0.158) &  &  &  &   \\ 
Future house prices correctly predicted (Ref. incorrect) &  & $-$0.017 &  &  &  \\ 
  &  & (0.105) &  &  &  \\ 
Expects that own income will over the next 10years...\\
 $\quad$ increase less than prices  (Ref. more or equal) &  & 0.178 &  &  &   \\ 
  &  & (0.133) &  &  & \\ 
 $\quad$ increase the same as prices  &  & $-$0.048 &  &  &   \\ 
  &  & (0.157) &  &  &   \\ 
Tenure status (Ref. renter)\\
 $\quad$ own house w/o mortgage  &  & 0.186$^{*}$ &  &  &   \\ 
  &  & (0.110) &  &  & \\ 
 $\quad$ own house with mortgage &  & 0.046 &  &  & \\ 
  &  & (0.130) &  &  &  \\ 
Acquisition mode (Ref.: purchased)\\
 $\quad$ house gifted  &  &  & 0.226 &  &   \\ 
  &  &  & (0.234) &  &   \\ 
 $\quad$ house inherited  &  &  & 0.218 &  &   \\ 
  &  &  & (0.190) &  &  \\ 
 $\quad$ other  &  &  & $-$0.292 &  &  \\ 
  &  &  & (0.344) &  &  \\ 
Owns other real estate = yes &  &  & 0.200 &  &   \\ 
  &  &  & (0.141) &  &   \\ 
Was involved in housing transactions in the past = No  &  &  & 0.051 &  &  \\ 
  &  &  & (0.129) &  &   \\ 
Future plans (Ref.: owner wanting to become a renter)\\
 is home owner, does not want to become home renter &  &  &  & $-$0.080 &  \\ 
  &  &  &  & (0.163) &   \\ 
Future plans (Ref.: renter wanting to become an owner)\\
 is renter, does not want to become home owner  &  &  &  &  & $-$0.048   \\ 
  &  &  &  &  & (0.162)   \\ 
Constant & $-$2.112$^{***}$ & $-$2.042$^{***}$ & $-$1.941$^{***}$ & $-$1.902$^{***}$ & $-$1.789$^{***}$  \\ 
  & (0.289) & (0.136) & (0.098) & (0.125) & (0.148)  \\ 
 \midrule \\ 
Observations & 3,735 & 3,708 & 2,254 & 1,437 & 2,312 \\ 
Log Likelihood & $-$1,443.840 & $-$1,455.523 & $-$899.870 & $-$540.082 & $-$929.528  \\ 
AIC & 2,907.679 & 2,923.045 & 1,811.739 & 1,084.163 & 1,863.056  \\ 
\bottomrule
\bottomrule
\end{tabular}} 
\end{center}
\begin{footnotesize}
\textit{Note:} The Table reports estimation results for probit models regressing the probability of a correct answer to Question 2 (UMP) on sets of control variables: respondents' socio-economic characteristics, general financial literacy, and previous housing market experience. $^{*}$p$<$0.1; $^{**}$p$<$0.05; $^{***}$p$<$0.01. 
\end{footnotesize}
\end{table}

%%%%%%%%%%%%%%%%%%%%%
\section{The Assumed Connection Between Monetary Policy and Housing Markets: With and Without Expert Guidance}\label{sec:beliefs}
In this section, we elicit participants’ ex-ante understanding of the relationship between housing market developments and monetary policy, and examine how ``monetary policy literacy'' relates to broader measures of financial and economic literacy. We further investigate whether expert guidance can help align public perceptions with economic theory.

%%%%%%%%%%%%%%%%
\subsection{The Association between Interest Rates and House Prices}\label{sec:interlinkage_hp_interest}

We start by asking directly whether participants suspect any kind of relationship between interest rates charged by credit institutes and house prices. These questions are aimed to transmit the notion of a specific aspect of inflation and the outcome of the monetary transmission process -- mortgage rates payable.

\begin{snugshade*}
\begin{question}
(HM 1)\label{HM.1}\\
``There is a connection between interest rates I have to pay to credit institutes when taking out a mortgage, and the price of houses/apartments.'' \\
Disagree / \textbf{Agree}
%    \item No.

\end{question}
\end{snugshade*}

Overall, Question \autoref{HM.1} reveals that the vast majority of respondents (75.72\%, that is 1,746 in counts) indeed thinks that there is such kind of nexus, whereas 24.28\% (560) reject this.

Subsequently, we ask about the suspected \emph{direction} of this connection. In other words, we elicit whether people assume a direct (co-movement) or indirect (anti-movement) relationship. To reduce ambiguity and capture consistent responses, the question was presented in two separate parts (Question \autoref{HMD.1} and Question \autoref{HMD.2}).

\begin{snugshade*}
\begin{question}
(HMD 1)\label{HMD.1}\\
``When the interest rate decreases, house prices decrease too.''\\
\textbf{Disagree} / Agree
\end{question}
\end{snugshade*}

\begin{snugshade*}
\begin{question}
(HMD 2)\label{HMD.2}\\
``When the interest rate decreases, house prices rise.''\\ Disagree / \textbf{Agree}
\end{question}
\end{snugshade*}

\begin{figure}[H]
\centering
\caption{Questions \autoref{HMD.1} and \autoref{HMD.2}: House price reactions to interest rate decreases.}
\label{fig:connection}
\begin{subfigure}{.5\textwidth}
  \centering
    \caption{Direct relationship (``When the interest rates decreases, house prices decrease too.'')}
  \includegraphics[width=0.99\linewidth]{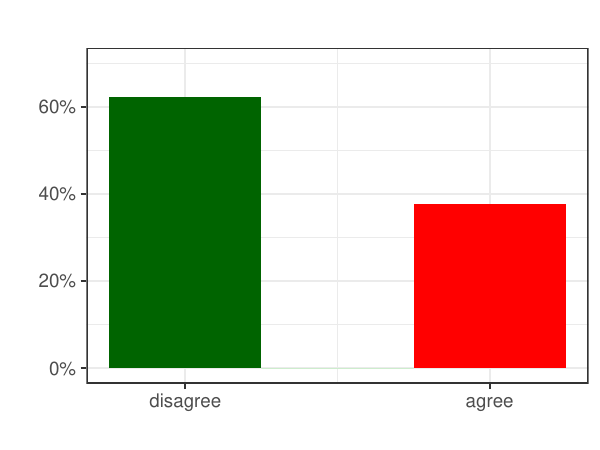}
  \label{fig:fig_connection_direct}
\end{subfigure}%
\begin{subfigure}{.5\textwidth}
  \centering
 \caption{Indirect relationship (``When the interest rate decreases, house prices rise'')}
  \includegraphics[width=0.99\linewidth]{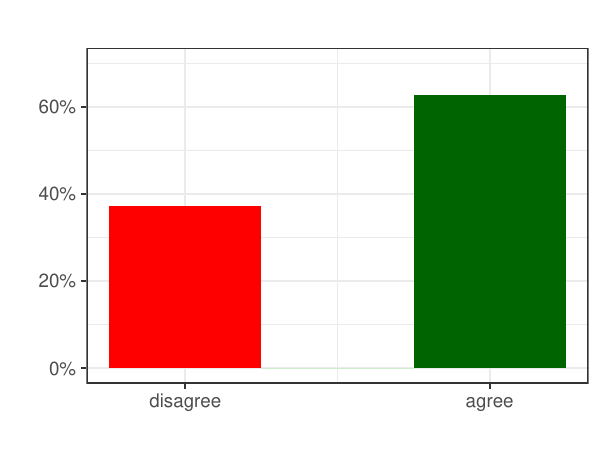}
  \label{fig:fig_connection_indirect}
\end{subfigure}
    \begin{footnotesize}
    \emph{Notes:} The Figures show answers to the mirrored Questions \autoref{HMD.1} and \autoref{HMD.2}. Full results are reported in \autoref{fig:connection}.
\end{footnotesize}
\end{figure}

The responses are quite symmetric confirming internal reliability of our results: economic In-line with economic reasoning \citep{hedlund2017monetary}, \autoref{fig:fig_connection_direct} shows that 61\% of respondents disagree with the notion that decreases in interest rate were associated with falling house price. \autoref{fig:fig_connection_indirect} indicates that, independently of the first question, 63\% also agree with the reciprocal statement, i.e., that decreases in the interest rates will lead to increasing (nominal) house price.

To understand the entire response patterns, we compile a contingency table to simultaneously assess both CMP questions for the same people (\autoref{tab:directindirect}). We identify a share of 46\% answering correctly to both questions and 21\% answering wrongly both times. The remaining respondents get at least one of the two wrong. Thus, these participants have a sound understanding of the relationship of interest rates and house price inflation. Note that these are less than half of the survey participants indicating severe knowledge gaps among the adult population in the countries assessed.

\begin{table}[H]
    \begin{center}
            \caption{Contingency Table: CMP (HMD 1 and HMD 2)}
    \label{tab:directindirect}
    \begin{tabular}{ll c c}
    \toprule 
\toprule 
& & \multicolumn{2}{c}{HMD 2} \\
&        & disagree & \textbf{agree} \\ 
        \cline{3-4}
       \multirow{4}{*}{HMD 1} &
        \multicolumn{1}{c|}{\textbf{disagree}} & 594  & \textbf{1,754} \\
         &  \multicolumn{1}{c|}{}& (16\%) & \textbf{(47\%)} \\
       & \multicolumn{1}{c|}{agree} &  807    & 612 \\
         &  \multicolumn{1}{c|}{ }& (21\%) & (16\%) \\       
        \bottomrule
        \bottomrule
    \end{tabular}
        \end{center}
    \begin{footnotesize}
    \emph{Notes:} The contingency table jointly assesses answers to Questions \autoref{HMD.1} and \autoref{HMD.2}. Correct answers are marked in bold. Overall probabilities are reported.
\end{footnotesize}
\end{table}

%%%%%%%%%%%%%%
\subsection{The Role of Monetary Policy Communication}\label{sec:communication}

%%%%%%%%%%%%%
\subsubsection{Set-up}\label{sec:setup}

Communication can shape expectations. 
Central bank communication indeed influences financial markets \citep{swanson2021measuring} and information about inflation can shape households' future inflation expectations \citep{coibion2022monetary, hoffmann2021effects}.
Central bank communication thus aims to make use of this and has accordingly been a rapidly growing area of research \citep[see][for a survey of this literature]{ehrmann2025trust}.

While central banks' interest in shaping inflation expectations is comprehensible, communication with the broader public may influence other -- potentially unintended -- aspects of monetary policy perceptions as well. Can information about the monetary policy-house price nexus shape households' stated beliefs? 

\cite{dolls2025factors} studies how information provision can alter policy views about rent control in Germany. They show that new information is not blindly incorporated into people's opinions, yet altering of views largely depends on prior beliefs, which affect perceived credibility and political neutrality of the received information. The mere provision of information may therefore not be sufficient to effectively alter policy views.
Accordingly, we do not simply provide new information in a neutral way, yet make use of expert profiles as credible source of information.

In our study, we introduce a randomised information treatment into our study and test whether experts as communicators in general help to alter beliefs and whether the type of expert plays a crucial role. 
Thus, we repeat questions \ref{HMD.1} and \ref{HMD.2}, but this time also provide information on experts' opinions on a nexus between housing prices and interest rates. The aim of this information treatment is to provide participants with some extra guidance on how this nexus is generally seen by experts and to study their willingness to trust them. 

We differentiate between a central bank (question \ref{CBT.1}) and an research (question \ref{RT.1}) economist providing this information by quoting statements publicly made by a representative of the Taiwan central bank and a research economist at the University of Toronto, respectively. Selecting these two quotes was driven by the fact that they clearly state a \emph{direction} of the connection between house prices and interest rates. Inline with economic theory, both economists argue in favour of an \emph{indirect} relationship between interest rates and housing prices, namely that if interest rates decrease, house prices will most likely increase.

\begin{snugshade*}
\begin{question}
(Info Central Banker)\label{CBT.1}\\
``There are central bankers arguing that when interest rates fall, house prices will rise. For example, Deputy Governor Chen Nan-kuang of Taiwan’s central bank states that ``loose monetary policy [note: meaning low interest rates] is indeed one of the main reasons for rising house prices.'' (Source: The Taiwan Banker NO.145\footnotemark)
\\
In the light of this claim, we will present you the last two statements again. Do you rather agree or disagree?" \\
\end{question}
\end{snugshade*}

\footnotetext{\cite{Chen2022_HousingPricesFinancialStability}.}

\begin{snugshade*}
\begin{question}
(Info Researcher)\label{RT.1}\\
"There are researchers arguing that when interest rates fall, house prices will rise. For example, Ozkan at the University of Toronto and his coauthors state that a ``reduction in the interest rate reduces the cost of borrowing, alleviates credit constraints and increases the demand for housing. The increase in demand for housing increases real house prices.''
(Source: Ozkan et al. 2017\footnotemark)
\\
In the light of this claim, we will present you the last two statements again. Do you rather agree or disagree?" \\
\end{question}
\end{snugshade*}

\footnotetext{\cite{hedlund2017monetary}.}

Every participant was shown only one of the two statements. The allocation to the two groups was randomised. 
After having been shown this information, participants were again asked to choose between two statements mentioning two opposing directions how house prices and interest rates could be related to each other.

This set-up allows us to assess the updating of beliefs upon provision of different types of experts' opinions by means of assessing switching behaviour between their initially chosen answer and the answer chosen upon receiving additional information.

%%%%%%%%%
\subsubsection{The General Role of Expert' Opinions to Shape Beliefs}\label{sec:general_expert_info}

\begin{figure}[h]
\caption{Switching Behaviour}
\centering
\begin{subfigure}{.48\textwidth}
  \centering
    \caption{Direct relationship: interest rate decreases, house prices decrease}
  \includegraphics[width=0.9\linewidth]{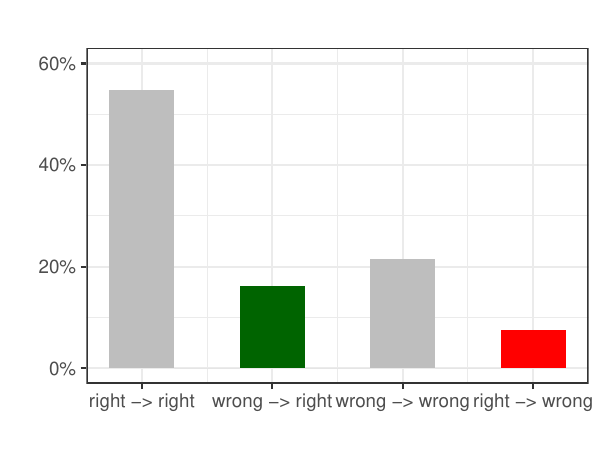}
  \label{fig:fig_connection_direct_switching}
\end{subfigure}%
\hfill
\begin{subfigure}{.48\textwidth}
  \centering
    \caption{Indirect relationship: interest rate decreases, house prices rise}

  \includegraphics[width=0.9\linewidth]{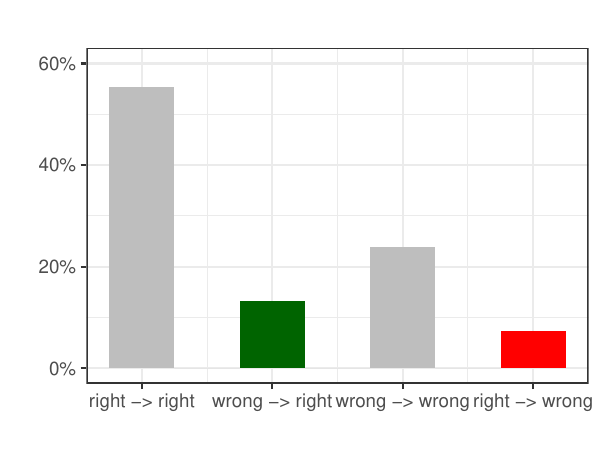}
  \label{fig:fig_connection_indirect_switching}
\end{subfigure}
\label{fig:fig_switch_total}
\end{figure}
\autoref{fig:fig_switch_total} shows the respondents' aggregate switching behaviour merging responses from participants receiving additional information by any type of expert. 
First, a vast majority of participants that had originally selected a correct answer, stick to this choice after being presented with additional information further affirming them in their choice.

We find that the majority of respondents does not change their prior selected answer in-line with a majority having had selected the correct answer also in the first place: 53\% (Question 4) and 54\% (Question 5), respectively, do not alter their correct responses after having received extra information. Conditional on selecting the correct answer, 88\% stick to it after the information treatment.

Second, about $13\%\text{–}16\%$ of respondents are ``good switchers,'' i.e., they move from an initially ``wrong'' to the ``correct'' answer. These respondents appear to have been convinced by the experts' statements. While the information treatment does not motivate the majority of initially incorrect respondents to switch, a sizeable share still does so ($36\%\text{–}43\%$).

A small minority ($7\%$ in total) initially answered correctly but switched to the wrong answer after the information treatment. We label these participants as ``bad switchers.''

This finding confirms that communication can improve the understanding of a sizeable share of respondents who begin with incorrect perceptions. Moreover, the clear majority of participants, namely $71\%/69\%$, did not contradict the experts' guidance. This group includes both respondents who originally answered correctly and maintained their answer after receiving information, as well as those who switched from an initially wrong to a correct answer. Taken together with the presence of a sizeable share of ``good switchers,'' this provides additional evidence for the notion that a majority of participants appear generally willing to accept expert guidance in this domain.

 % 608/(608+810)
 % 498/(498+903)
 
Zooming in, \autoref{tab:switch_country} in the Appendix reports the country-specific results.
There is substantial heterogeneity across countries both for the direct and indirect questions. 
In general, the lowest share of participants actively contradicting experts (i.e., switching from an initially correct to an incorrect answer upon information) is lowest in Sweden and the UK for both questions.
To assess the other extreme, i.e., the share of respondents following experts' advise yields a very similar picture across countries: Again, the share of respondents who initially chose the wrong answer and -- after the information treatment -- switched to the correct one ranges between $33\%$ and $37\%$ for the direct question in most countries, but reaches as high as $69\%$ in Sweden.%
\footnote{This number results from: $258/(258+115)=0.69$.}
Switching rates for the indirect question are lower across all countries yet reach the highest values in the UK with $45\%$ and again Sweden with $39\%$ versus $32\%$-$34\%$ for the other countries.%
\footnote{These number results from $68/(68+107)=0.39$ for Sweden and $93/(93+112)$ for the UK.})

Thus, respondents in Sweden and to a certain degree also the UK appear most likely to revise their opinions in accordance with the experts' statements. This high level of trust in experts aligns with findings from the EUROBAROMETER \citep{european2021european}, which measures general trust in institutions, professions, and groups. Sweden records the highest level of trust in scientists among all participating countries, followed by the UK: $84\%$ of Swedes and $79\%$ of Brits agree that scientists working at universities or government-funded research organisations are best qualified to explain the societal impact of scientific and technological developments. Support for this question is much lower in Germany (59\%), Austria (46\%) and Italy (62\%).

By contrast, respondents in Germany are most likely to revise their opinion in the direction opposite to what the expert statements suggest: $17\%$ of those who initially answered correctly switch to an incorrect answer after receiving the information treatment. Germans are indeed among the most sceptical in Europe regarding trust in experts, with only $59\%$ agreeing with the EUROBAROMETER statement above.

In the remaining countries, the share of respondents switching away from the experts' position is comparatively small, ranging from $8\%$ in the UK to $15\%$ in Italy.

%%%%%%%%%%%%%
\subsubsection{Does It Matter Who provides Experts Advice?}\label{sec:type_expert}

\begin{figure}[h]
\caption{Switching Behaviour by Source of Information}
\centering
\begin{subfigure}{.45\textwidth}
  \centering
    \caption{interest rate decreases, house prices decrease}
  \includegraphics[width=0.9\linewidth]{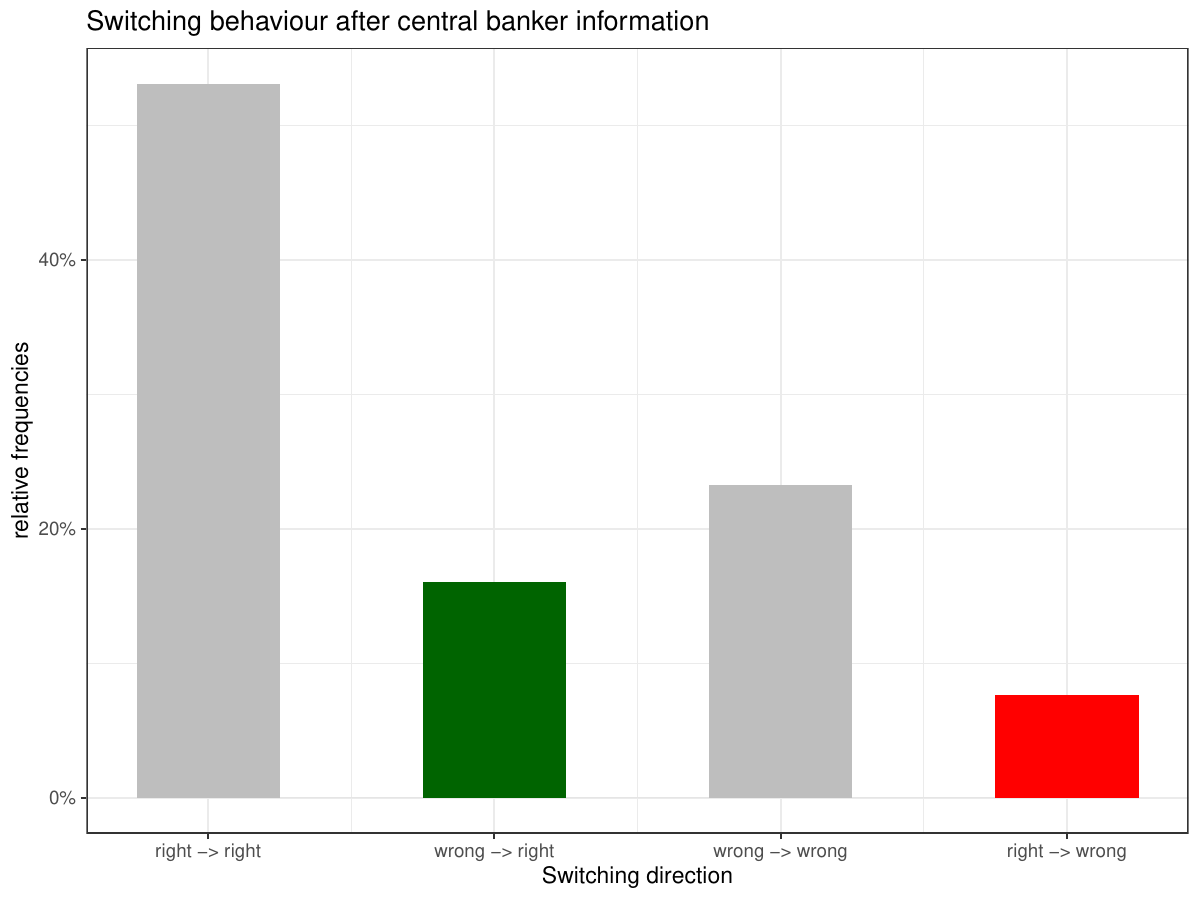}
  \captionsetup{font={footnotesize}}
  \label{fig:fig_switch_cb_direct}
\end{subfigure}%
\begin{subfigure}{.45\textwidth}
  \centering
    \caption{interest rate decreases, house prices rise}
  \includegraphics[width=0.9\linewidth]{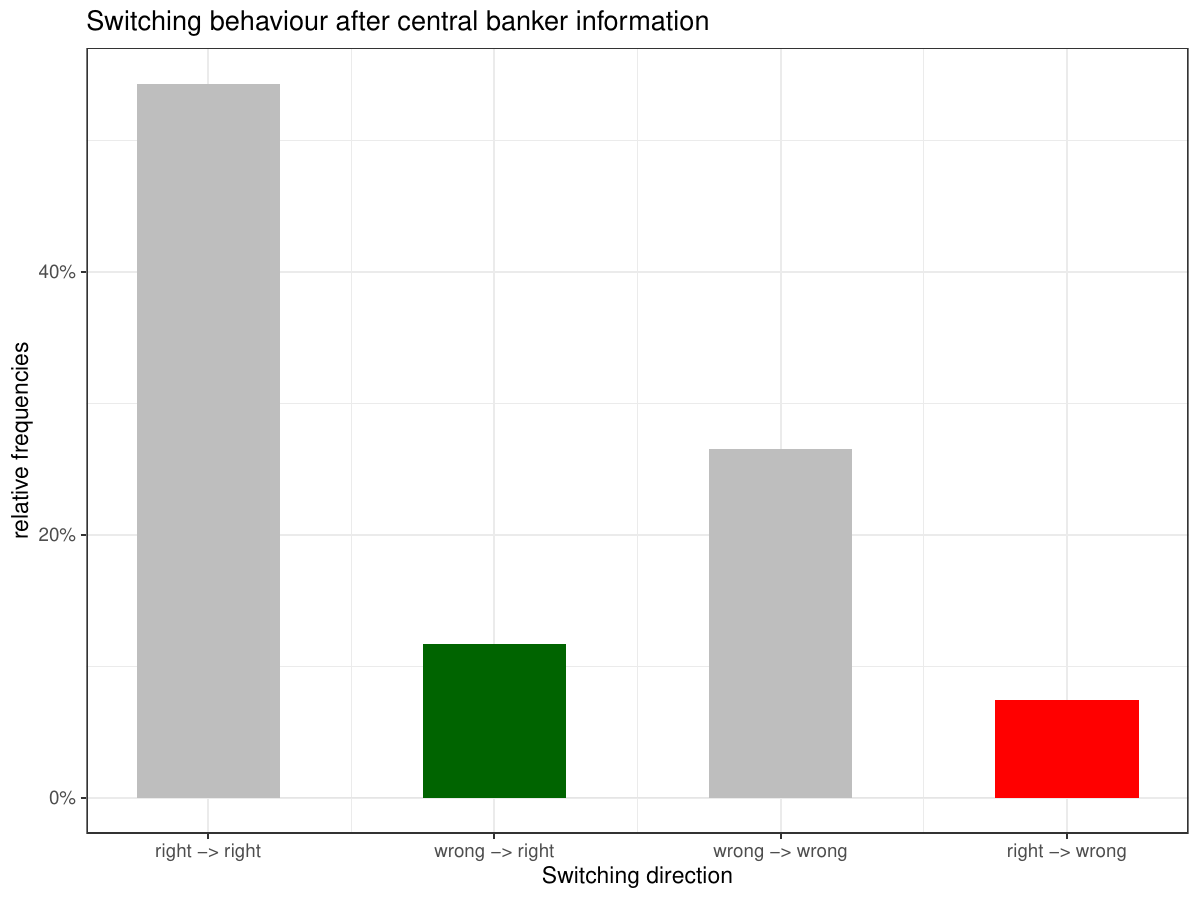}
    \captionsetup{font={footnotesize}}
  \label{fig:fig_switch_cb_indirect}
\end{subfigure}
\begin{subfigure}{.45\textwidth}
  \centering
    \caption{interest rate decreases, house prices decrease}
  \includegraphics[width=0.9\linewidth]{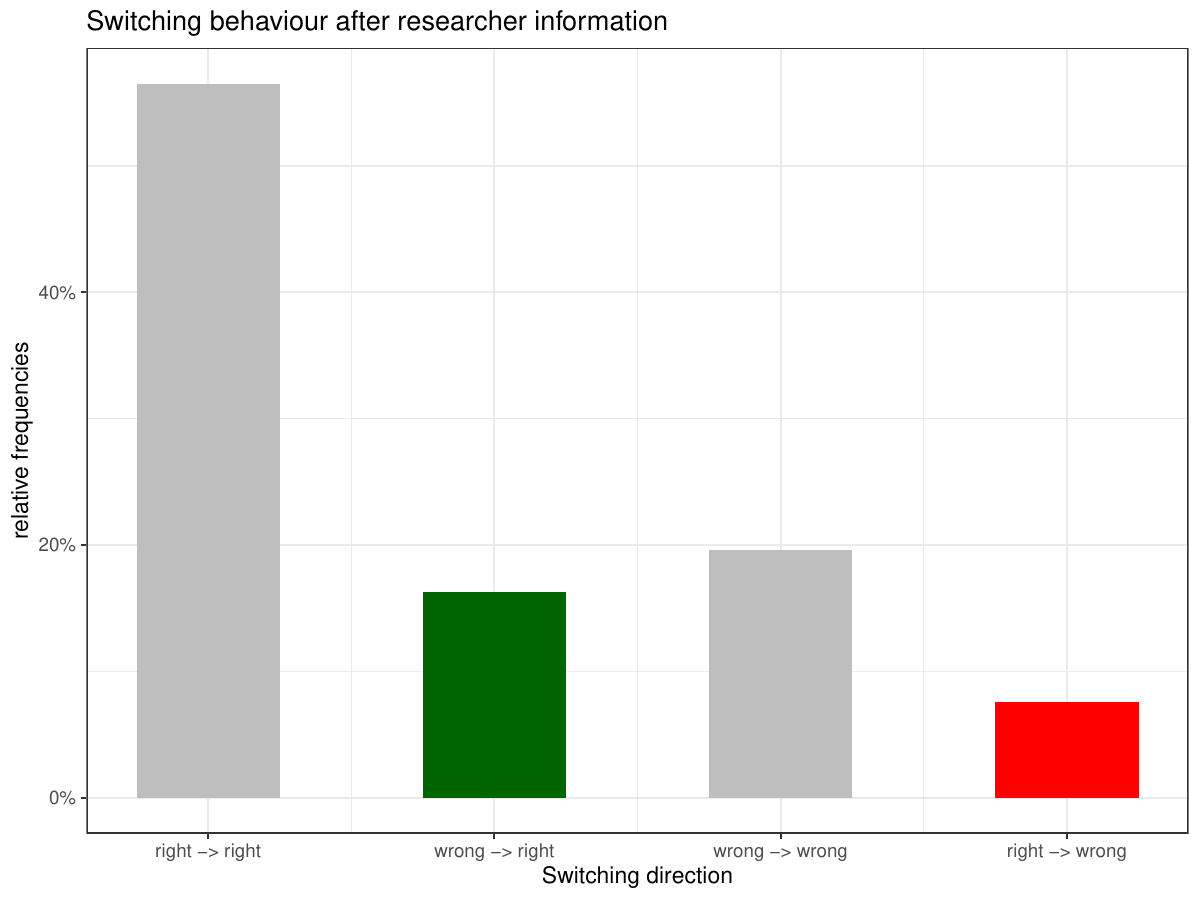}
    \captionsetup{font={footnotesize}}
  \label{fig:fig_switch_econ_direct}
\end{subfigure}%
\begin{subfigure}{.45\textwidth}
  \centering
    \caption{interest rate decreases, house prices rise}
  \includegraphics[width=0.9\linewidth]{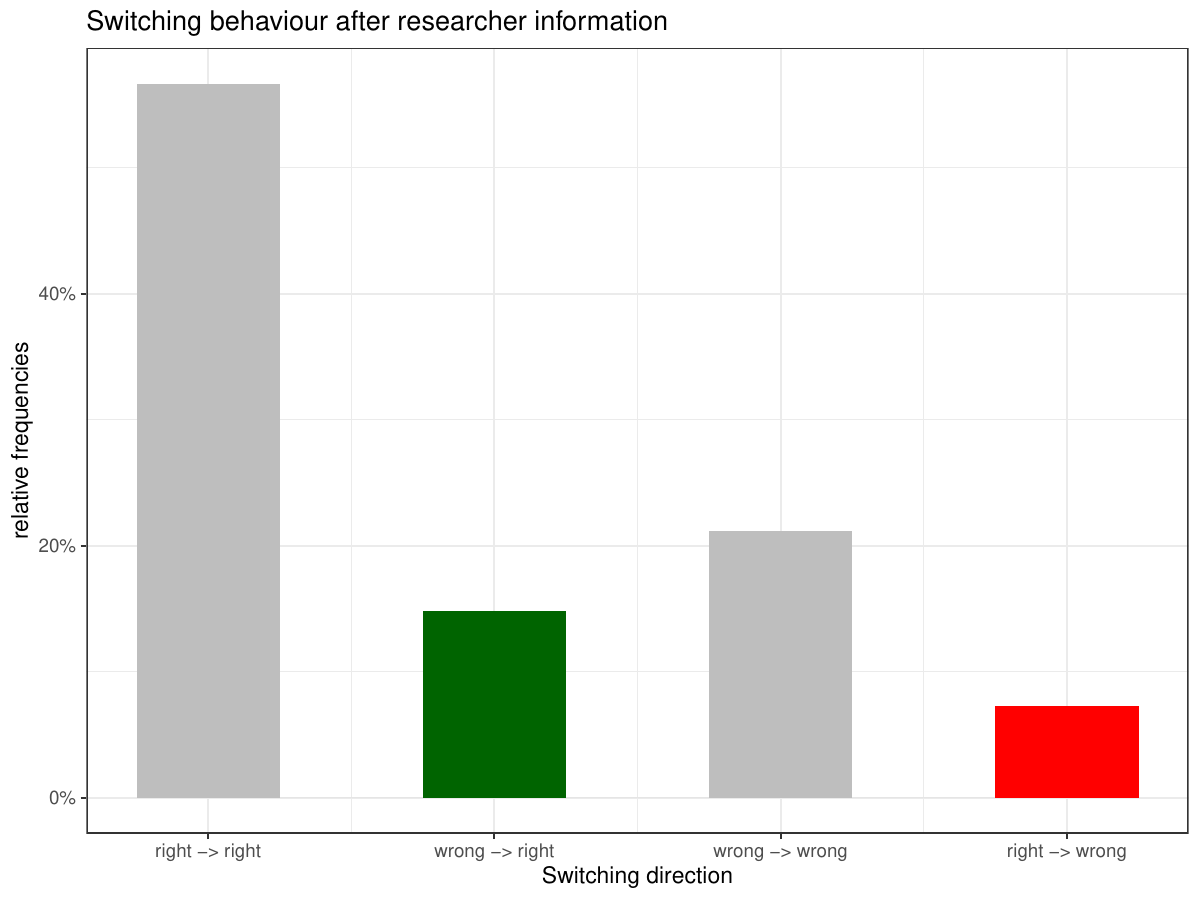}
    \captionsetup{font={footnotesize}}
  \label{fig:fig_switch_econ_indirect}
\end{subfigure}
\label{fig:fig_switch_info}
\end{figure}

Next, we assess the importance of the \emph{type of economist} providing expert advise. \autoref{fig:fig_switch_info} therefore splits the information effect by information provider: a central banker or an academic economist. We first focus on the group of participants updating their response from an initially wrong to a correct answer (``good switcher'').

While there is only little variation by expert statement for the question on a direct connection (Questions \autoref{HMD.1}: HMD 1), a substantial number of participants switch when stating an indirect relationship (Question \autoref{HMD.2}: HMD 2).
Central banker information triggers 12\% of respondents to correct their initially wrong answer, while researcher information animates 15\% to do so. We run a test on equal proportions \citep{newcombe1998interval} to judge whether this detected difference by information provider is statistically significantly different from zero and indeed, the test rejects the null hypothesis of equal proportions ($\chi^2=7.7705$, $p$-value$=0.0053$).

Although initial responses to both statements are almost perfectly symmetric (see \autoref{fig:connection}), incorrect answers are corrected more frequently for the direct question (HMD 1) than for the indirect one (HMD 2).

The share of ``bad switchers,'' i.e., those switching from an initially correct answer to a wrong one is roughly the same across the type of information giver as well as indirect vs. direct relationship. Indeed, we cannot reject the null of equal properties ($\chi^2 = 0.0094$, $p$-value = $0.9227$).

\begin{table}[!htbp] 
  \caption{The Role of Experts' Communication: Ordered Logit Models} 
  \label{tab:ologit} 
  \begin{center}
  \resizebox{0.7\textwidth}{!}{%
\begin{tabular}{@{\extracolsep{5pt}}lcccc} 
\toprule
\toprule
 & \multicolumn{4}{c}{\textit{Dependent variable:}} \\
 & \multicolumn{4}{c}{Incorrigibles, Susceptibles, Experts} \\ 
\cmidrule{2-5} 
& \multicolumn{2}{c}{HMD.1 (direct)} & \multicolumn{2}{c}{HMD.2 (indirect)} \\
\cmidrule{2-5}
\\
\multicolumn{5}{l}{\textbf{Panel A:} Full Sample}\\
\\
 & (1) & (2) & (3) & (4)\\ 
\cmidrule{2-5}  
 Academic Economist Info & 0.171$^{**}$ & 0.170$^{**}$ & 0.162$^{**}$ & 0.181$^{***}$ \\ 
  & (0.067) & (0.068) & (0.067) & (0.069) \\  
\cmidrule{2-5} 
SE &  & \checkmark & & \checkmark\\
\cmidrule{2-5}
Observations & 3,479 & 3,454 & 3,490 & 3,465 \\ 
Log Likelihood & -3316.872 & -3196.444 & -3259.627 & -3118.192 \\
AIC & 6639.744 & 6416.888 & 6525.253  & 6260.383  \\
\midrule\\

\multicolumn{5}{l}{\textbf{Panel B:} Restricted Sample -- Consistent Answers}\\
\\ & (5) & (6) & (7) & (8)\\ 
\cmidrule{2-5}
 Academic Economist Info & 0.192$^{**}$ & 0.208$^{**}$ & 0.207$^{**}$ &  0.224$^{**}$ \\ 
  & (0.085) & (0.088) & (0.085) & (0.088) \\ 
 \cmidrule{2-5}
 SE &  &  \checkmark & & \checkmark\\
 \cmidrule{2-5}
Observations & 2,373 & 2,363 & 2,391 & 2,381 \\ 
Log Likelihood & -2041.704 & -1953.628 & -2055.059 & -1970.428 \\
AIC & 4089.408 & 3931.256 & 4116.118 &  3964.856 \\
\midrule\\

\multicolumn{5}{l}{\textbf{Panel C:} Restricted Sample -- Understands CMP}\\
\\ & (9) & (10) & (11) & (12)\\ 
\cmidrule{2-5}
 Academic Economist Info & 0.311$^{***}$ & 0.281$^{***}$ & 0.165$^{*}$ & 0.194$^{*}$  \\ 
  & (0.097) & (0.099) & (0.097) & (0.101) \\ 
 \cmidrule{2-5}
 SE &  &  \checkmark & & \checkmark\\
\cmidrule{2-5}
Observations & 1,726 & 1,715 & 1,751 & 1,740 \\ 
Log Likelihood & -1576.023 & -1525.935 & -1553.338 &  -1468.719 \\
AIC & 3158.047 & 3075.87 & 3112.675 & 2961.438 \\
\bottomrule
\bottomrule
\end{tabular}
}
\end{center}
\begin{footnotesize}
\textit{Note:} The Table reports Odds Ratios. The full sample includes all participants, while the restricted sample filter likely more trustworthy participants: Panel B 
includes only participants that respond in a consistent manner to the two mirrored questions testing for MP literacy (Questions \autoref{HMD.1} and \autoref{HMD.2}), and Panel C includes only participants to have demonstrated basic understanding of conventional monetary policy (Question \autoref{CMP.1}).
SE denotes the same set of socio-economic characteristics as used in Model (1) of \autoref{tab:hetero_CMP} and \autoref{tab:hetero_UMP}. {$^{*}$p$<$0.1; $^{**}$p$<$0.05; $^{***}$p$<$0.01} 
\end{footnotesize}
\end{table}

To assess the robustness of these descriptive results, we examine the effect of communicator type in a multivariate framework. Specifically, we classify participants by their performance to construct an ordinal outcome variable and estimate ordered logit models.
Concretely, we categorize respondents as follows: those who remain incorrect even after receiving additional information (``The Incorrigibles''), those who change their incorrect answer to the correct one (``The Susceptibles''), and those who are consistently correct (``The Experts''). 
\autoref{tab:ologit} reports the results for both questions. Panel A shows results for all participants. 
Results are clear: Information provided by an academic economist is more successful for updating beliefs than information provided by a central banker. The effects are consistently robust to the inclusion of socio-economic characteristics. When constructing the estimation samples in a way to filter more trustworthy responses, effect sizes are consistently even larger. 
Precisely, Panel B filters answers of participants that had filled the survey with more dedication, and hence restricts the sample to those participants that answered in a consistent way to the two mirrored (direct and indirect) questions on interest rates and house prices (Questions \autoref{HMD.1} and \autoref{HMD.2}). This consistency requirement means improved reliability.
Panel C filters answers of participants that had a better general MP understanding, and hence limits the sample to those participants that correctly answered the conventional monetary policy question (Question \autoref{CMP.1}). 
Effect sizes are systematically larger in magnitude in the restricted samples; indicating that filtering participants fulfilling certain minimum criteria (having consistently answered mirrored questions or having demonstrated basic monetary policy literacy) potentially selects those filling out the survey in a more serious manner, suggesting that the true effect size is likely larger than the one measured in Panel A.

We further test whether there are heterogeneities in this tendency to follow advise by academic economists more or whether this is a general finding. 
%We, therefore, stepwise interacted the randomized information treatment dummy with participants' characteristics and stated perceptions. 
Concretely, we first make use of stated trust in the government and institutions,%
\footnote{\emph{(S1) How good do you think the government and tax authorities are at things like efficient use of resources, avoidance of mistakes and preventing fraud?}} 
perceptions of corruption,%
\footnote{\emph{(S2) Please tell me whether you think the government and tax authorities in your country give special advantages to certain people or deal with everyone equally?}}
and political participation in the form of voting turnout which has been shown to correlate with trust in policy and politics \citep{Hooghe2017TrustAndElections} in general.
Further, we assess heterogeneity by education attainment (tertiary vs. less), and expectation or past receipt of an inheritance%
\footnote{\emph{(S3) Do you or any other member of your household expect to receive an inheritance or a substantial gift (worth more than GBP 8,500), including money or any other assets (e.g., houses, land, cars, etc.) in the future?}; \emph{(S4) Have you or any member of your household ever received an inheritance or a substantial gift, including money or any other major assets (e.g., houses, land, cars, etc.)?}}.

We split all dimensions -- that are not binary in nature -- into two categories (high/low) to sustain sufficient observations per group, and interact the binary outcome with the type of information provider.
\autoref{tab:heterogeneity1} reports results for the direct question (Question \autoref{HMD.1}),  and \autoref{tab:heterogeneity2} for the indirect question (Question \autoref{HMD.2}).

%For each heterogeneity test, we split observations into two groups based on the answer to these questions and interact the resulting binary variables with the type of information provider.

%Concretely, we split samples by having high or low trust in government%
%\footnote{\emph{(S1) How good do you think the government and tax authorities are at things like efficient use of resources, avoidance of mistakes and preventing fraud?}}, think about the government as corrupt or not\footnote{\emph{(S2) Please tell me whether you think the government and tax authorities in your country give special advantages to certain people or deal with everyone equally?}}, voted in the last election\footnote{As in previous models}, have high or low education\footnote{As in previous models, tertiary or lower}, expect an inheritance\footnote{\emph{(S3) Do you or any other member of your household expect to receive an inheritance or a substantial gift (worth more than GBP 8,500), including money or any other assets (e.g., houses, land, cars, etc.) in the future?}} or inherited in the past\footnote{\emph{(S4) Have you or any member of your household ever received an inheritance or a substantial gift, including money or any other major assets (e.g., houses, land, cars, etc.)?}}. We split all these dimensions that are not binary in nature into two categories to sustain sufficient observations per group. \autoref{tab:heterogeneity1} reports results for the direct question (Question \autoref{HMD.1}), \autoref{tab:heterogeneity2} for Question \autoref{HMD.2}.

For no single dimension, the interaction term is significantly different from zero, no matter whether we include controls or not, and, hence, we conclude that there is no systematic deviation among the dimensions tested from the general result that consumers are more likely to follow academic economists' advise in this setting.

Overall, these results indicate that information conveyed by central bankers is somewhat less effective at aligning public perceptions with economic theory than information provided by research economists. Moreover, the absence of heterogeneous treatment effects suggests that this pattern holds broadly across different groups, underscoring the generality of the finding.

While further research is needed to assess the robustness of these results across countries, over time, and with different types of expert communicators, our findings provide an initial indication that monetary policy communication may be most effective when it draws on a strategic mix of experts, including academic economists who can emphasize and clarify the underlying theoretical foundations.

%%%%%%%%%%%%%%%%%%%%%%%%%%%
\section{Conclusions}\label{sec:conclusions}

This paper studies an important but largely unexplored dimension of the monetary policy–housing nexus: how it is perceived by the general public. While the transmission of interest rates to housing markets is well established in theory and extensively documented in empirical research, little is known about whether households understand this mechanism, how strongly they believe in it, and how their beliefs respond to expert information. By combining a large-scale, cross-country survey-experiment with an information treatment, we provide some of the first systematic evidence on how monetary policy actions are linked to house price expectations in the minds of the general public.

Our results show that the public's understanding of monetary policy is highly uneven across policy instruments. A majority of respondents demonstrate a basic grasp of conventional monetary policy, particularly the link between inflation and policy rate adjustments. In contrast, literacy regarding unconventional monetary policy remains very low, highlighting a substantial information gap. This asymmetry suggests that experience-based learning plays a crucial role: conventional monetary policy is more easily understood because its effects are more visible in everyday financial decisions, especially in mortgage markets, whereas unconventional tools remain abstract and distant from households’ lived experiences.

At the same time, we find strong intuitive awareness of the housing channel of monetary policy. Three quarters of respondents perceive a clear connection between interest rates and house prices, and most correctly associate falling interest rates with rising house prices. This indicates that, even if formal knowledge of monetary policy instruments is limited, households possess a coherent mental model of how monetary policy affects housing markets. Given the central role of housing in household wealth and financial stability, this perception is economically meaningful and likely to shape both expectations and behaviour.

A central contribution of our study lies in demonstrating that beliefs about the monetary policy–housing nexus are malleable. A non-negligible share of respondents revise incorrect prior beliefs when presented with concise, credible information from experts. This finding reinforces the growing evidence that expectations are not fixed, but responsive to targeted communication. It also underscores the potential power of well-designed central bank and academic outreach in improving public understanding of monetary policy transmission.

Our results further reveal important heterogeneity in how expert information is received. Participants respond more strongly to statements by academic economists than to those by central bank economists. This suggests a credibility gap that is highly relevant for monetary authorities. While central banks are currently the primary communicators of monetary policy, they may not always be perceived as the most trusted source of economic expertise by the general public. This finding calls for a rethinking of communication strategies, including closer cooperation with academic institutions and a stronger emphasis on transparency and independence in messaging.

We also show that monetary policy literacy is closely linked to socio-economic characteristics and life experiences. Education, gender, age, and direct involvement in housing and mortgage markets all shape individuals' understanding of monetary policy. Conventional monetary policy literacy follows a life-cycle pattern, peaking around the age when housing purchases typically occur, while unconventional monetary policy literacy does not exhibit such experiential learning. The persistent gender gap in both forms of literacy, even after accounting for age and education, highlights the need for more inclusive approaches to economic communication and education.

Moreover, the association between lower literacy and indicators of economic insecurity and political disengagement points to a broader socio-economic dimension of monetary policy understanding. Households who feel economically vulnerable or less connected to civic institutions appear to be systematically less informed about monetary policy. This suggests that improving monetary policy literacy is not only a matter of information provision, but also closely related to issues of trust, inclusion, and social participation.

Taken together, our findings have several implications. First, they highlight that the housing market is a key channel through which households interpret and experience monetary policy. Second, they show that expectations about house prices are deeply intertwined with perceptions of monetary policy actions. Third, they underline the importance of credible, accessible, and differentiated communication strategies that reach beyond financial experts and engage the broader public.

By documenting how people perceive the monetary policy–housing nexus, how they update their beliefs, and whom they trust as sources of expertise, this study contributes to a more comprehensive understanding of expectation formation. Ultimately, the effectiveness of monetary policy depends not only on its technical design, but also on how well it is understood and trusted by the public. Improving monetary policy literacy -- especially in relation to housing markets -- can therefore be seen as an integral component of a well-functioning monetary policy framework.

%%%%%%%%%%%%%
\section{Declaration of Generative AI and AI-assisted technologies in the Manuscript Preparation Process}

During the preparation of this work the authors used \emph{Writefull} and \emph{ChatGTP} in order to enhance writing, as well as \emph{DeepL} for translation. After using this tool, the authors reviewed and edited the content as needed and take full responsibility for the content of the published article.
%%%%%%%%%%%%%%%%%%%%%%%%%%%%%%%%%%%%%%%%%%%%%%%%%%%% 
%%% Literature
%%%%%%%%%%%%%%%%%%%%%%%%%%%%%%%%%%%%%%%%%%%%%%%%%%%%

%\newpage
\bibliographystyle{apalike}
\bibliography{bibliography}

%%%%%%%%%%%%%%%%%%%%%%%%%%%%%%%%%%%%%%%%%%%%%%%%%%%% 
%%% Appendix
%%%%%%%%%%%%%%%%%%%%%%%%%%%%%%%%%%%%%%%%%%%%%%%%%%%%

\newpage
\appendix

\section*{Appendix}\label{app:tables}

\begin{table}[h!]
 \caption{Summary Statistics: Participants' Socioeconomic Characteristics}  \label{tab:summary_all}
\begin{center}   
\resizebox{0.8\textwidth}{!}{
\begin{tabular}{rllrrr}
  \toprule
  \toprule
\multicolumn{2}{l}{Variable} & Value & Frequency & Percent & Cum. Percent \\ 
  \midrule
\multicolumn{2}{l}{Gender}   & female & 1996 & 51.67 & 51.67 \\ 

 &  & male & 1755 & 45.43 & 97.10 \\ 
   &  & other & 14 & 0.36 & 97.46 \\ 
   &  & NA & 98 & 2.54 & 100.00 \\ 
      \midrule
\multicolumn{2}{l}{Marital Status}  & single & 1013 & 26.22 & 26.22 \\ 
   &  & partnership/married & 2433 & 62.98 & 89.20 \\ 
   &  & divorced/widowed & 319 & 8.26 & 97.46 \\ 
   &  & NA & 98 & 2.54 & 100.00 \\ 
         \midrule
\multicolumn{4}{l}{Have one or more children} \\
   && yes & 2021 & 52.32 & 52.32 \\ 
   &  & no & 1743 & 45.12 & 97.44 \\ 
   &  & NA & 99 & 2.56 & 100.00 \\ 
     \midrule

\multicolumn{4}{l}{Subjective self-assessed probability of job loss (within 6 months)} \\
   & low & 1 & 1726 & 44.68 & 44.68 \\ 
   &  & 2 & 503 & 13.02 & 61.79 \\ 
   & & 3 & 302 & 7.82 & 69.61 \\ 
   &  & 4 & 185 & 4.79 & 74.40 \\ 
   &  & 5 & 237 & 6.14 & 80.54 \\ 
   &  & 6 & 268 & 6.94 & 87.48 \\ 
   &  & 7 & 155 & 4.01 & 91.49 \\ 
   &  & 8 & 153 & 3.96 & 95.45 \\ 
   &  & 9 & 79 & 2.05 & 97.50 \\ 
   & high  & 10 & 158 & 4.09 & 48.77 \\ 
   &  & NA & 97 & 2.51 & 100.00 \\ 
    \midrule

\multicolumn{4}{l}{Degree of life satisfaction} \\
   & low & 1 & 124 & 3.21 & 3.21 \\ 
  &  & 2 & 145 & 3.75 & 13.22 \\ 
   &  & 3 & 225 & 5.82 & 19.04 \\ 
   & & 4 & 268 & 6.94 & 25.98 \\ 
   &  & 5 & 391 & 10.12 & 36.10 \\ 
   &  & 6 & 513 & 13.28 & 49.38 \\ 
   & & 7 & 671 & 17.37 & 66.75 \\ 
   &  & 8 & 784 & 20.30 & 87.05 \\ 
   & & 9 & 404 & 10.46 & 97.51 \\ 
     & high & 10 & 242 & 6.26 & 9.47 \\ 
        &  & NA & 96 & 2.49 & 100.00 \\ 
       \midrule
\multicolumn{4}{l}{Highest level of education attained}\\
   &  & below tertiary & 2451 & 63.45 & 63.45 \\ 
   &  & tertiary & 1314 & 34.02 & 97.47 \\ 
   &  & NA & 98 & 2.54 & 100.00 \\ 
          \midrule
\multicolumn{4}{l}{Did vote in the last election eligible}\\
   &  & Yes & 3434 & 88.89 & 88.89 \\ 
   &  & No & 427 & 11.05 & 99.94 \\ 
   &  & NA & 2 & 0.05 & 100.00 \\ 
   \midrule
\multicolumn{4}{l}{Self-assessed position in the country's earnings distribution}\\
   & lowest quintile & 1 & 753 & 19.49 & 19.49 \\ 
   & & 2 & 753 & 19.49 & 38.98 \\ 
   &  & 3 & 752 & 19.47 & 58.45 \\ 
   &  & 4 & 752 & 19.47 & 77.92 \\ 
   & highest quintile & 5 & 752 & 19.47 & 97.39 \\ 
   &  & NA & 101 & 2.61 & 100.00 \\ 
  \bottomrule
   \bottomrule
\end{tabular}}\\
\end{center}
\footnotesize
\emph{Notes:} The Table reports overall socio-economic characteristics across all five countries.
\end{table}

\begin{table}[h]
\centering
\caption{Pearson's $\chi^2$ test and Cramér's V for CMP and UMP by country}
\label{tab:chi_cramer_country}
\begin{tabular}{lrrrr}
\toprule \toprule
Country & $\chi^2$ & df & $p$-value & Cramér's V \\
\midrule
Austria   & 1.503 & 1 & 0.220     & 0.044 \\
Germany   & 5.459 & 1 & 0.019$^{**}$ & 0.088 \\
Italy     & 6.657 & 1 & 0.010$^{**}$ & 0.091 \\
Sweden    & 0.024 & 1 & 0.876     & 0.005 \\
UK        & 1.347 & 1 & 0.246     & 0.046 \\
\bottomrule \bottomrule
\end{tabular}
\begin{minipage}{0.9\textwidth}
\footnotesize{\emph{Notes:} Cramér's V ranges within $[0;1]$.
Significance levels: $^{*} p<0.05$, $^{**} p<0.01$, $^{***} p<0.001$.}
\end{minipage}
\end{table}

\begin{figure}[H]
    \begin{center}
    \caption{Answers to Question 1 (CMP) and Question 2 (UMP)}\label{fig:replicate_Q1_Q2_counts}
    \begin{subfigure}{.45\textwidth}
  \centering
            \includegraphics[width=1\textwidth]{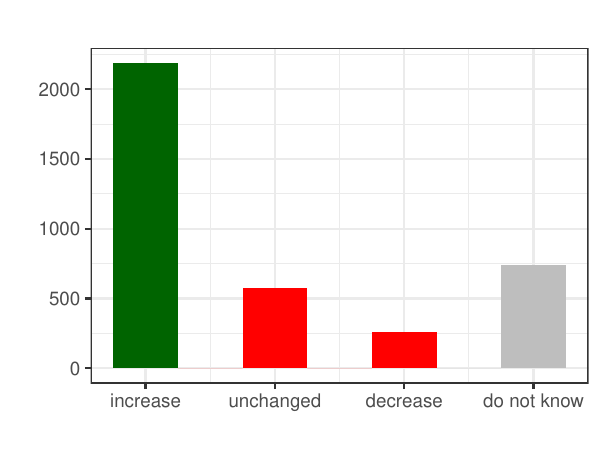}
            \caption{CMP}    
    \label{fig:fig_ECB_Q1_counts}
    \end{subfigure}\hfill
    \begin{subfigure}{.45\textwidth}
  \centering
    \includegraphics[width=1\textwidth]{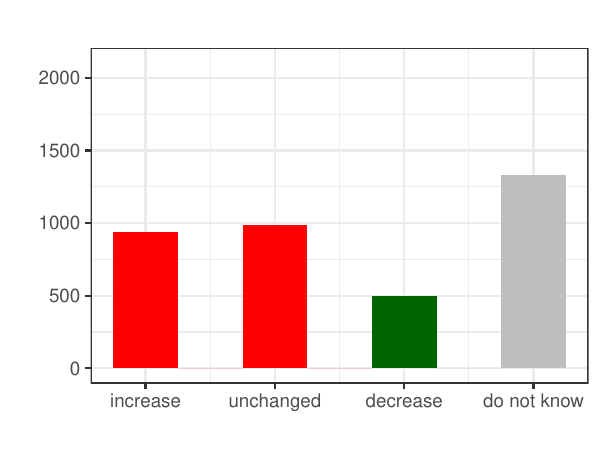}
        \caption{UMP}
    \label{fig:fig_ECB_Q2_counts}
    \end{subfigure}
    \end{center}
    \begin{footnotesize}
        \emph{Notes:} The chart shows answers in counts to questions eliciting the connection between inflation and CMP (\autoref{fig:fig_ECB_Q1_counts}) and UMP (\autoref{fig:fig_ECB_Q2_counts}), respectively. %\autoref{fig:fig_ECB_checker_counts} shows the intersection of both questions. 
        Correct answers are marked in green.
    \end{footnotesize}
\end{figure}

\begin{table}[!htbp]  
  \caption{Determinants of MP Literacy (Extended Socioeconomic Controls)} 
  \label{tab:hetero_MP_all} 
  \begin{center}
  \resizebox{0.65\textwidth}{!}{    
\begin{tabular}{@{\extracolsep{5pt}}lcccc} 
\toprule
\toprule

 & \multicolumn{2}{c}{CMP} & \multicolumn{2}{c}{UMP} \\ 
\cmidrule{2-3} \cmidrule{4-5}
 & \multicolumn{1}{c}{Unweighted} & \multicolumn{1}{c}{Weighted} 
 & \multicolumn{1}{c}{Unweighted} & \multicolumn{1}{c}{Weighted} \\ 
\\
 & (1) & (2) & (3) & (4) \\ 
\midrule \\ 

Gender (Ref.: female)\\
 $\quad$ male & 0.215$^{**}$ & 0.220$^{**}$ & 0.328$^{**}$ & 0.319$^{**}$ \\ 
  & (0.095) & (0.104) & (0.141) & (0.156) \\ 

Education attainment (Ref.: low)\\
 $\quad$ high education & 0.299$^{**}$ & 0.295$^{*}$ & 0.439$^{**}$ & 0.453$^{*}$ \\ 
  & (0.144) & (0.158) & (0.215) & (0.236) \\ 

Age\\
  & 0.038$^{*}$ & 0.039 & $-$0.017 & $-$0.018 \\ 
  & (0.023) & (0.025) & (0.031) & (0.035) \\ 

Age$^2$\\
  & $-$0.0003 & $-$0.0003 & 0.0003 & 0.0003 \\ 
  & (0.0003) & (0.0003) & (0.0003) & (0.0004) \\ 

Marital status (Ref.: single)\\
 $\quad$ married/partnered & $-$0.044 & $-$0.044 & $-$0.049 & $-$0.053 \\ 
  & (0.098) & (0.107) & (0.137) & (0.151) \\ 
 $\quad$ divorced/widowed & $-$0.031 & $-$0.037 & $-$0.177 & $-$0.198 \\ 
  & (0.157) & (0.171) & (0.232) & (0.255) \\ 

Children\\
  & 0.120 & 0.119 & 0.122 & 0.121 \\ 
  & (0.086) & (0.094) & (0.123) & (0.135) \\ 

Likelihood of job loss within 6 months\\
  & $-$0.049$^{***}$ & $-$0.050$^{**}$ & 0.026 & 0.029 \\ 
  & (0.018) & (0.020) & (0.025) & (0.028) \\ 

Life satisfaction\\
  & 0.016 & 0.015 & $-$0.048$^{*}$ & $-$0.046 \\ 
  & (0.018) & (0.020) & (0.026) & (0.029) \\ 

Country sample (Ref.: Austria)\\
 $\quad$ Germany & $-$0.679$^{***}$ & $-$0.680$^{***}$ & $-$0.182 & $-$0.185 \\ 
  & (0.120) & (0.130) & (0.167) & (0.182) \\ 
 $\quad$ Italy & $-$0.647$^{***}$ & $-$0.648$^{***}$ & 0.193 & 0.192 \\ 
  & (0.121) & (0.134) & (0.163) & (0.182) \\ 
 $\quad$ Sweden & $-$0.332$^{**}$ & $-$0.325$^{**}$ & $-$0.220 & $-$0.232 \\ 
  & (0.138) & (0.153) & (0.186) & (0.207) \\ 
 $\quad$ United Kingdom & $-$0.290$^{**}$ & $-$0.292$^{**}$ & $-$0.634$^{***}$ & $-$0.636$^{***}$ \\ 
  & (0.129) & (0.137) & (0.195) & (0.207) \\ 

Did not vote (Ref.: voted)\\
  & $-$0.370$^{***}$ & $-$0.365$^{***}$ & 0.098 & 0.087 \\ 
  & (0.122) & (0.134) & (0.176) & (0.194) \\ 

Employment status (Ref.: full-time employed)\\
 $\quad$ unemployed & $-$0.005 & $-$0.013 & 0.247 & 0.252 \\ 
  & (0.139) & (0.153) & (0.183) & (0.202) \\ 
 $\quad$ student & $-$0.061 & $-$0.082 & 0.058 & 0.068 \\ 
  & (0.151) & (0.167) & (0.216) & (0.240) \\ 
 $\quad$ retired/other & 0.134 & 0.139 & $-$0.034 & $-$0.015 \\ 
  & (0.147) & (0.161) & (0.210) & (0.230) \\ 

Income (Ref.: middle)\\
 $\quad$ low & $-$0.022 & $-$0.015 & $-$0.076 & $-$0.063 \\ 
  & (0.107) & (0.117) & (0.156) & (0.171) \\ 
 $\quad$ high & 0.041 & 0.033 & 0.120 & 0.140 \\ 
  & (0.117) & (0.127) & (0.163) & (0.178) \\ 

Area of residence (Ref.: urban)\\
 $\quad$ rural & $-$0.011 & $-$0.018 & $-$0.147 & $-$0.144 \\ 
  & (0.079) & (0.087) & (0.112) & (0.123) \\ 

Subjective social class (Ref.: middle)\\
 $\quad$ lower & 0.073 & 0.065 & 0.041 & 0.048 \\ 
  & (0.100) & (0.110) & (0.147) & (0.161) \\ 
 $\quad$ upper & $-$0.360$^{*}$ & $-$0.372 & 0.355 & 0.346 \\ 
  & (0.211) & (0.229) & (0.282) & (0.309) \\ 

Interactions\\
 $\quad$ male $\times$ high education & 0.121 & 0.119 & 0.200 & 0.189 \\ 
  & (0.158) & (0.172) & (0.221) & (0.242) \\ 
 $\quad$ high education $\times$ job loss & $-$0.038 & $-$0.035 & $-$0.069$^{*}$ & $-$0.070 \\ 
  & (0.029) & (0.032) & (0.042) & (0.045) \\ 

Constant & $-$0.294 & $-$0.308 & $-$1.868$^{**}$ & $-$1.852$^{**}$ \\ 
  & (0.597) & (0.657) & (0.834) & (0.919) \\ 

\midrule \\ 
Observations & 3,178 & 3,178 & 3,183 & 3,183 \\ 
Log Likelihood & $-$2,073.815 & $-$2,073.865 & $-$1,219.806 & $-$1,219.851 \\ 
AIC & 4,197.630 & 4,197.730 & 2,489.613 & 2,489.702 \\ 
\bottomrule
\bottomrule
\end{tabular}} 
\end{center}

\begin{footnotesize}
\textit{Note:} The Table reports probit estimation results for monetary policy literacy. Columns (1) and (2) refer to correct answers to the CMP question (unweighted and weighted), while columns (3) and (4) refer to the UMP question (unweighted and weighted). Robust standard errors are reported in parentheses. $^{*}$p$<$0.1; $^{**}$p$<$0.05; $^{***}$p$<$0.01.
\end{footnotesize}
\end{table}

\begin{table} 
  \caption{Determinants of CMP Literacy: Weighted} 
  \label{tab:hetero_CMPw} 
  \begin{center}
  \resizebox{0.9\textwidth}{!}{    
\begin{tabular}{@{\extracolsep{5pt}}lccccc} 
\toprule \toprule 
 & \multicolumn{5}{c}{\textit{Dependent variable: CMP correct}} \\ 
\cmidrule{2-6} 
\\ & \multicolumn{1}{c}{SE} &\multicolumn{1}{c}{Expec} & \multicolumn{3}{c}{Homewonership}  \\ 
\\ & (1) & (2) & (3) & (4) & (5) \\ 
\cmidrule{2-6}  
Gender (Ref.: female)\\
 $\quad$ male & 0.249$^{***}$ &  &  &  &  \\ 
  & (0.076) &  &  &  &   \\ 
Education attainment (Ref.: low)\\
 $\quad$ high education & 0.284$^{***}$ &  &  &  &  \\ 
  & (0.083) &  &  &  &   \\ 
Age & 0.015$^{***}$ &  &  &  &  \\ 
  & (0.003) &  &  &  &   \\ 
Likelihood of job loss within 6 months (10steps Likert scale) & $-$0.073$^{***}$ &  &  &  &   \\ 
  & (0.014) &  &  &  &   \\ 
Country Sample (Ref.: Austria)\\
 $\quad$ Germany & $-$0.700$^{***}$ &  &  &  &   \\ 
  & (0.120) &  &  &  &   \\ 
 $\quad$ Italy & $-$0.673$^{***}$ &  &  &  &  \\ 
  & (0.119) &  &  &  & \\ 
 $\quad$ Sweden & $-$0.242$^{**}$ &  &  &  &  \\ 
  & (0.121) &  &  &  &  \\ 
 $\quad$ United Kingdom & $-$0.309$^{**}$ &  &  &  &  \\ 
  & (0.124) &  &  &  &   \\ 
Did not vote  (Ref. did vote)& $-$0.407$^{***}$ &  &  &  &  \\ 
  & (0.119) &  &  &  &   \\ 
Future house prices correctly predicted (Ref. incorrect) &  & 0.012 &  &  &   \\ 
  &  & (0.080) &  &  &  \\ 
Expects that own income will over the next 10years...\\
 $\quad$ increase less than prices  (Ref. more or equal)&  & 0.340$^{***}$ &  &  &   \\ 
  &  & (0.098) &  &  &  \\ 
 $\quad$ increase the same as prices  &  & $-$0.008 &  &  &   \\ 
  &  & (0.112) &  &  &  \\ 
Tenure status (Ref. renter)\\
 $\quad$ own house w/o mortgage  &  & 0.134 &  &  &   \\ 
  &  & (0.083) &  &  &  \\ 
 $\quad$ own house with mortgage &  & 0.470$^{***}$ &  &  &   \\ 
  &  & (0.099) &  &  &  \\ 
Acquisition mode (Ref.: purchased)\\
 $\quad$ house gifted  &  &  & $-$0.243 &  &   \\ 
  &  &  & (0.192) &  &   \\ 
 $\quad$ house inherited  &  &  & $-$0.304$^{**}$ &  &   \\ 
  &  &  & (0.154) &  &  \\ 
 $\quad$ other  &  &  & 0.136 &  &  \\ 
  &  &  & (0.241) &  &  \\ 
Owns other real estate = yes &  &  & $-$0.077 &  &  \\ 
  &  &  & (0.114) &  &  \\ 
Was involved in housing transactions in the past = No  &  &  & $-$0.443$^{***}$ &  &   \\ 
  &  &  & (0.100) &  & \\ 
Future plans (Ref.: renter wanting to become an owner)\\
 Is renter, does not want to become home owner  &  &  &  & $-$0.060 &  \\ 
  &  &  &  & (0.119) &   \\ 
Future plans (Ref.: owner wanting to become a renter)\\
 is home owner, does not want to become home renter &  &  &  &  & 0.246$^{*}$  \\ 
  &  &  &  &  & (0.126)   \\ 
Constant & 0.532$^{**}$ & $-$0.017 & 0.714$^{***}$ & 0.215$^{**}$ & 0.222$^{*}$ \\ 
  & (0.220) & (0.098) & (0.076) & (0.093) & (0.115)  \\ 
 \midrule \\[-1.8ex] 
Observations & 3,729 & 3,702 & 2,252 & 1,433 & 2,310  \\ 
Log Likelihood & $-$2,436.627 & $-$2,487.743 & $-$1,491.641 & $-$987.149 & $-$1,548.562 \\ 
AIC & 4,893.253 & 4,987.485 & 2,995.283 & 1,978.298 & 3,101.124  \\ 
\bottomrule \bottomrule 
\end{tabular}} 
\end{center}
\begin{footnotesize}
    \emph{Notes:} The Table reports estimation results for weighted probit models regressing the probability of a correct answer to Question 1 (CMP) on sets of control variables: respondents' socio-economic characteristics, general financial literacy, and previous housing market experience. $^{*}$p$<$0.1; $^{**}$p$<$0.05; $^{***}$p$<$0.01. 
\end{footnotesize}
\end{table}

\begin{table}[!htbp] 
  \caption{Determinants of UMP Literacy: Weighted} 
  \label{tab:hetero_UMPw} 
  \begin{center}
  \resizebox{0.9\textwidth}{!}{    
\begin{tabular}{@{\extracolsep{5pt}}lccccc} 
\toprule\toprule 
 & \multicolumn{5}{c}{\textit{Dependent variable: UMP correct}} \\ 
\cmidrule{2-6} 
\\ & \multicolumn{1}{c}{SE} & \multicolumn{1}{c}{Expec} & \multicolumn{3}{c}{Homewonership}  \\ 
 & (1) & (2) & (3) & (4) & (5) \\ 
\cmidrule{2-6} 
Gender (Ref.: female)\\
 $\quad$ male & 0.412$^{***}$ &  &  &  &   \\ 
  & (0.109) &  &  &  &  \\ 
Education attainment (Ref.: low)\\
 $\quad$ high education & 0.334$^{***}$ &  &  &  &  \\ 
  & (0.116) &  &  &  &  \\ 
Age & $-$0.001 &  &  &  \\ 
  & (0.004) &  &  &  &  \\ 
Likelihood of job loss within 12 months (10steps Likert scale) & $-$0.001 &  &  &  &   \\ 
  & (0.020) &  &  &  &   \\ 
Country Sample (Ref.: Austria)\\
 $\quad$ Germany & $-$0.196 &  &  &  &   \\ 
  & (0.167) &  &  &  &   \\ 
 $\quad$ Italy & 0.130 &  &  &  &   \\ 
  & (0.160) &  &  &  &  \\ 
 $\quad$ Sweden & $-$0.171 &  &  &  & \\ 
  & (0.166) &  &  &  &  \\ 
 $\quad$ United Kingdom & $-$0.590$^{***}$ &  &  &  &  \\ 
  & (0.185) &  &  &  &  \\ 
Did not vote  (Ref. did vote) & 0.052 &  &  &  &  \\ 
  & (0.174) &  &  &  &   \\ 
Future house prices correctly predicted (Ref. incorrect) &  & $-$0.020 &  &  &   \\ 
  &  & (0.115) &  &  &  \\ 
Expects that own income will over the next 10years...\\
 $\quad$ increase less than prices  (Ref. more or equal) &  & 0.183 &  &  &   \\ 
  &  & (0.146) &  &  &  \\ 
 $\quad$ increase the same as prices  &  & $-$0.050 &  &  &  \\ 
  &  & (0.172) &  &  &  \\ 
Tenure status (Ref. renter)\\
 $\quad$ own house w/o mortgage  &  & 0.177 &  &  &   \\ 
  &  & (0.121) &  &  &  \\ 
 $\quad$ own house with mortgage &  & 0.034 &  &  &   \\ 
  &  & (0.143) &  &  &  \\ 
Acquisition mode (Ref.: purchased)\\
 $\quad$ house gifted  &  &  & 0.245 &  &   \\ 
  &  &  & (0.258) &  &  \\ 
 $\quad$ house inherited  &  &  & 0.239 &  &  \\ 
  &  &  & (0.210) &  &   \\ 
 $\quad$ other  &  &  & $-$0.256 &  &   \\ 
  &  &  & (0.379) &  &   \\ 
Owns other real estate = yes &  &  & 0.207 &  &  \\ 
  &  &  & (0.155) &  &  \\ 
Was involved in housing transactions in the past = No  &  &  & 0.041 &  &  \\ 
  &  &  & (0.142) &  &   \\ 
Future plans (Ref.: owner wanting to become a renter)\\
 is home owner, does not want to become home renter &  &  &  & $-$0.085 &    \\ 
  &  &  &  & (0.179) &    \\ 
Future plans (Ref.: renter wanting to become an owner)\\
 is renter, does not want to become home owner  &  &  &  &  & $-$0.038   \\ 
  &  &  &  &  & (0.180)   \\ 
Constant & $-$2.098$^{***}$ & $-$2.045$^{***}$ & $-$1.957$^{***}$ & $-$1.901$^{***}$ & $-$1.811$^{***}$  \\ 
  & (0.318) & (0.149) & (0.107) & (0.138) & (0.164) \\ 
 \midrule
Observations & 3,735 & 3,708 & 2,254 & 1,437 & 2,312  \\ 
Log Likelihood & $-$1,443.854 & $-$1,455.550 & $-$899.909 & $-$540.083 & $-$929.558  \\ 
AIC & 2,907.707 & 2,923.099 & 1,811.818 & 1,084.165 & 1,863.117 \\ 
\bottomrule
\bottomrule 
\end{tabular}} 
\end{center}
\begin{footnotesize}
    \emph{Notes:} The Table reports estimation results for probit models (survey-weighted) regressing the probability of a correct answer to Question 2 (UMP) on sets of control variables: respondents' socio-economic characteristics, general financial literacy, and previous housing market experience. $^{*}$p$<$0.1; $^{**}$p$<$0.05; $^{***}$p$<$0.01. 
\end{footnotesize}
\end{table}

\begin{table}[!htbp]
\caption{Switching Behaviour After Information Treatment, by Country}
\label{tab:switch_country}
\begin{center}
\small
\setlength{\tabcolsep}{4pt}
\begin{tabular}{l l rr rr}
\toprule
\toprule
& & \multicolumn{2}{c}{Direct Question} & \multicolumn{2}{c}{Indirect Question} \\
\cmidrule(lr){3-4}\cmidrule(lr){5-6}
Country & Switching Direction
& $n$ & \% & $n$ & \% \\
\midrule
Austria
& right $\rightarrow$ right & 493 & 0.631 & 380 & 0.487 \\
& wrong $\rightarrow$ right &  77 & 0.099 & 111 & 0.142 \\
& wrong $\rightarrow$ wrong & 157 & 0.201 & 235 & 0.301 \\
& right $\rightarrow$ wrong &  53 & 0.068 &  54 & 0.069 \\
& NA                    &   1 & 0.001 &   1 & 0.001 \\
\addlinespace
Germany
& right $\rightarrow$ right & 375 & 0.508 & 367 & 0.497 \\
& wrong $\rightarrow$ right &  86 & 0.117 &  96 & 0.130 \\
& wrong $\rightarrow$ wrong & 178 & 0.241 & 188 & 0.255 \\
& right $\rightarrow$ wrong &  75 & 0.102 &  63 & 0.085 \\
& NA                    &  24 & 0.033 &  24 & 0.033 \\
\addlinespace
Italy
& right $\rightarrow$ right & 408 & 0.499 & 349 & 0.427 \\
& wrong $\rightarrow$ right & 119 & 0.145 & 130 & 0.159 \\
& wrong $\rightarrow$ wrong & 210 & 0.257 & 261 & 0.319 \\
& right $\rightarrow$ wrong &  74 & 0.090 &  72 & 0.088 \\
& NA                    &   7 & 0.009 &   6 & 0.007 \\
\addlinespace
Sweden
& right $\rightarrow$ right & 402 & 0.481 & 604 & 0.722 \\
& wrong $\rightarrow$ right & 258 & 0.309 &  68 & 0.081 \\
& wrong $\rightarrow$ wrong & 115 & 0.138 & 107 & 0.128 \\
& right $\rightarrow$ wrong &  53 & 0.063 &  49 & 0.059 \\
& NA                    &   8 & 0.010 &   8 & 0.010 \\
\addlinespace
UK
& right $\rightarrow$ right & 383 & 0.557 & 387 & 0.562 \\
& wrong $\rightarrow$ right &  68 & 0.099 &  93 & 0.135 \\
& wrong $\rightarrow$ wrong & 148 & 0.215 & 112 & 0.163 \\
& right $\rightarrow$ wrong &  32 & 0.047 &  40 & 0.058 \\
& NA                    &  57 & 0.083 &  56 & 0.081 \\
\bottomrule
\bottomrule
\end{tabular}
\end{center}
\footnotesize
\emph{Notes:} The Table reports country-specific updating behaviour upon the provision of information by experts.
``Direct'' refers to Question \autoref{HMD.1}, HMD.1, and ``Indirect'' to Question \autoref{HMD.2}, HMD.2.
\end{table}

\begin{table}[!htbp] 
  \caption{The Effect of Age on CMP and UMP Literacy} 
  \label{tab:age} 
  \begin{center}
     % \resizebox{0.9\textwidth}{!}{%

\begin{tabular}{@{\extracolsep{5pt}}lccc} 
\toprule\toprule
 & \multicolumn{3}{c}{\textit{Dependent variable: CMP correct}} \\ 
\cline{2-4} 
\\ & (1) & (2) & (3) \\ 
\cmidrule{2-4} \\ 
  Age & 0.014$^{***}$ & 0.035$^{*}$ &   \\ 
  & (0.003) & (0.019) &\\ 
  Age$^2$ &  & $-$0.0002 &    \\ 
  &  & (0.0002) &    \\ 
  $\sqrt{\text{Age}}$ &  &  & 0.187$^{***}$  \\ 
  &  &  & (0.037)  \\ 
\midrule
SE & \checkmark & \checkmark & \checkmark\\
 
 \midrule 
Observations & 3,729 & 3,729 & 3,729  \\ 
Log Likelihood & $-$2,436.608 & $-$2,435.940 & $-$2,436.188 \\ 
AIC & 4,893.216 & 4,893.881 & 4,892.375  \\ 
\\
\midrule\\
 & \multicolumn{3}{c}{\textit{Dependent variable: UMP correct}} \\ 
\cline{2-4} 
\\ & (4) & (5) & (6) \\ 
\cmidrule{2-4} \\ 
 Age & $-$0.001 & $-$0.040 &   \\ 
  & (0.004) & (0.025) &   \\ 
   Age$^2$  &  & 0.0004 &    \\ 
  &  & (0.0003) &    \\ 
 $\sqrt{\text{Age}}$ &  &  & $-$0.018  \\ 
  &  &  & (0.052)  \\
  \midrule
SE & \checkmark & \checkmark & \checkmark\\
 
  \midrule
Observations & 3,735 & 3,735 & 3,735 \\ 
Log Likelihood & $-$1,443.840 & $-$1,442.607 & $-$1,443.798\\ 
AIC & 2,907.679 & 2,907.215 & 2,907.595 \\ 
\bottomrule\bottomrule

\end{tabular}
%}
\end{center}
\footnotesize
\emph{Notes:}
SE refers to the same set as used in Model (1) in \autoref{tab:hetero_CMP} and \autoref{tab:hetero_UMP}.
\end{table}

\begin{table}[!htbp] 
\caption{Heterogeneity Analysis -- Direct Effect} 
\label{tab:heterogeneity1} 
\begin{center}
\resizebox{0.95\textwidth}{!}{
\begin{tabular}{l cc c cc c cc c}
\toprule\toprule
 & \multicolumn{8}{c}{\textit{Dependent variables: Incorrigibles, Susceptibles, Experts}}\\
\midrule
 & \multicolumn{2}{c}{Trust} && 
   \multicolumn{2}{c}{Corruption} && 
   \multicolumn{2}{c}{Vote} \\
\cmidrule(lr){2-3}\cmidrule(lr){5-6}\cmidrule(lr){8-9}
 & (1) & (2) && (3) & (4) && (5) & (6) \\
\cmidrule(lr){2-3}\cmidrule(lr){5-6}\cmidrule(lr){8-9}
Academic Economist Info 
 & \num{0.177} & 0.233* && \num{0.234}*** & 0.232*** && 0.481* & 0.516** \\
 & (\num{0.124}) & (0.128) && (0.019) & (0.087) && (0.246) & (0.252) \\
\addlinespace
Group $\times$ Academic Economist Info 
 & \num{-0.007} & -0.091 && -0.206 & -0.391* && -0.279 & -0.310 \\
 & (\num{0.227}) & (0.151) && (0.213) & (0.223) && (0.213) & (0.218) \\
\midrule
SE &  & \checkmark &&  & \checkmark &&  & \checkmark \\
\midrule
Observations 
 & 3,729 & 3,453 && 3,729 & 3,454 && 3,729 & 3,454 \\
Log Likelihood 
 & -3,315.796 & -3,195.920 && -3,242.800 & -3,153.071 && -3,314.906 & -3,195.428 \\
AIC 
 & 4,893.216 & 6,419.8 && 4,893.881 & 6,334.1 && 4,892.375 & 6,416.9 \\
\midrule\midrule
 & \multicolumn{2}{c}{Education} && 
   \multicolumn{2}{c}{Expected inheritance} && 
   \multicolumn{2}{c}{Past inheritance} \\
\cmidrule(lr){2-3}\cmidrule(lr){5-6}\cmidrule(lr){8-9}
 & (1) & (2) && (3) & (4) && (5) & (6) \\
\cmidrule(lr){2-3}\cmidrule(lr){5-6}\cmidrule(lr){8-9}
Academic Economist Info 
 & 0.109 & 0.249 && 0.427 & 0.438 && 0.355 & 0.306 \\
 & (0.201) & (0.206) && (0.270) & (0.206) && (0.249) & (0.255) \\
\addlinespace
Group $\times$ Academic Economist Info 
 & 0.048 & -0.058 && -0.145 & -0.156 && -0.107 & -0.083 \\
 & (0.141) & (0.144) && (0.151) & (0.154) && (0.143) & (0.146) \\
\midrule
SE &  & \checkmark &&  & \checkmark &&  & \checkmark \\
\midrule
Observations 
 & 3,467 & 3,454 && 3,463 & 3,445 && 3,463 & 3,445 \\
Log Likelihood 
 & -3,302.202 & -3,196.363 && -3,299.094 & -3,189.868 && -3,296.536 & -3,189.211 \\
AIC 
 & 6,614.4 & 6,418.7 && 6,608.2 & 6,407.7 && 6,603.1 & 6,406.4 \\
\bottomrule\bottomrule
\end{tabular}
}
\end{center}
\footnotesize
\emph{Notes:} Ordered Logit Models on the outcome in Question \autoref{HMD.1}. 
SE refers to the same set as used in Model (1) in \autoref{tab:hetero_CMP} and \autoref{tab:hetero_UMP}.
\end{table}

\begin{table}[!htbp] 
\caption{Heterogeneity Analysis -- Indirect Effect} 
\label{tab:heterogeneity2} 
\begin{center}
\resizebox{0.95\textwidth}{!}{
\begin{tabular}{l cc c cc c cc c}
\toprule\toprule
 & \multicolumn{8}{c}{\textit{Dependent variables: Incorrigibles, Susceptibles, Experts}}\\
\midrule
 & \multicolumn{2}{c}{Trust} && 
   \multicolumn{2}{c}{Corruption} && 
   \multicolumn{2}{c}{Vote} \\
\cmidrule(lr){2-3}\cmidrule(lr){5-6}\cmidrule(lr){8-9}
 & (1) & (2) && (3) & (4) && (5) & (6) \\
\cmidrule(lr){2-3}\cmidrule(lr){5-6}\cmidrule(lr){8-9}
Academic Economist Info 
 & \num{0.132} & 0.137 && 0.233*** & 0.226*** && 0.289 & 0.377 \\
 & (0.126) & (0.082) && (0.082) & (0.084) && (0.241) & (0.248) \\
\addlinespace
Group $\times$ Academic Economist Info 
 & 0.040 & 0.060 && -0.220 & -0.138 && -0.114 & -0.175 \\
 & (0.149) & (0.153) && (0.143) & (0.147) && (0.207) & (0.212) \\
\midrule
SE &  & \checkmark &&  & \checkmark &&  & \checkmark \\
\midrule
Observations 
 & 3,488 & 3,464 && 3,489 & 3,465 && 3,488 & 3,465 \\
Log Likelihood 
 & -3,257.630 & -3,117.796 && -3,257.139 & -3,117.709 && -3,248.483 & -3,117.853 \\
AIC 
 & 6,525.3 & 6,263.6 && 6,524.3 & 6,263.4 && 6,507.0 & 6,261.7 \\
\midrule\midrule
 & \multicolumn{2}{c}{Education} && 
   \multicolumn{2}{c}{Expected inheritance} && 
   \multicolumn{2}{c}{Past inheritance} \\
\cmidrule(lr){2-3}\cmidrule(lr){5-6}\cmidrule(lr){8-9}
 & (1) & (2) && (3) & (4) && (5) & (6) \\
\cmidrule(lr){2-3}\cmidrule(lr){5-6}\cmidrule(lr){8-9}
Academic Economist Info 
 & -0.065 & -0.007 && 0.171 & 0.243 && -0.040 & 0.056 \\
 & (0.203) & (0.208) && (0.281) & (0.288) && (0.252) & (0.260) \\
\addlinespace
Group $\times$ Academic Economist Info  
 & 0.187 & 0.143 && -0.003 & -0.037 && 0.119 & 0.071 \\
 & (0.146) & (0.149) && (0.156) & (0.160) && (0.144) & (0.149) \\
\midrule
SE &  & \checkmark &&  & \checkmark &&  & \checkmark \\
\midrule
Observations 
 & 3,478 & 3,465 && 3,474 & 3,456 && 3,474 & 3,456 \\
Log Likelihood
 & -3,224.417 & -3,117.730 && -3,233.376 & -3,103.513 && -3,241.555 & -3,109.541 \\
AIC 
 & 6,458.8 & 6,261.5 && 6,476.8 & 6,235.0 && 6,493.1 & 6,247.1 \\
\bottomrule\bottomrule
\end{tabular}
}
\end{center}
\footnotesize
\emph{Notes:} Ordered Logit Models on the outcome in Question \autoref{HMD.2}. 
SE refers to the same set as used in Model (1) in \autoref{tab:hetero_CMP} and \autoref{tab:hetero_UMP}.
\end{table}

\end{document}